\documentclass[preprint,preprintnumbers,amsmath,amssymb]{revtex4-2}
\usepackage{graphicx}
\usepackage{graphics}
\usepackage{chngcntr}
\usepackage{bm}
\usepackage{epsfig}
\begin{document}
	
	\title{Exclusive nonleptonic $B_c$-meson decays to S-wave charmonium states}

	\author{Lopamudra Nayak$^{1}$\footnote{email address:lopalmn95@gmail.com}, P. C. Dash$^{1}$, Susmita Kar$^{2}$, N. Barik$^{3}$}
	\affiliation{$^{1}$ Department of Physics, Siksha $'O'$ Anusandhan (Deemed to be University), Bhubaneswar-751030,\\$^{2}$ Department of Physics, Maharaja Sriram Chandra Bhanja Deo University, Baripada-757003,\\ 
		$^{3}$ Department of Physics, Utkal University, Bhubaneswar-751004.}
	\begin{abstract}
	We study the exclusive two-body nonleptonic $B_c\to X_{c\bar{c}}M$ decays, where $X_{c\bar{c}}$ is either a ground(1S) or a radially excited (2S or 3S) charmonium and $M$ is a pseudoscalar $(P)$ or a vector $(V)$ meson. We consider here three categories of decays: $B_c \to PP, PV, VP$ decays within the framework of relativistic independent quark(RIQ) model based on a flavor-independent interaction potential in scalar-vector harmonic form. Using the factorization approximation, we calculate the weak form factors from the overlapping integrals of meson wave functions obtained in the RIQ model and predict the branching fractions for a set of exclusive nonleptonic $B_c$-decays in reasonable agreement with other model predictions. Some of the decays of interest are found to have branching fractions $\sim (10^{-3}-10^{-4})$ within the detection ability of the current experiments and can be precisely measured at LHCb in near future. In the wake of the recent measurement of $B_c\to J/\psi\pi(K)$, $B_c\to J/\psi\pi(D_s)$, $B_c\to \pi(J/\psi,\psi(2S))$ and $B_c\to J/\psi(\pi,\mu \nu)$ reported by the LHCb Collaborations, we predict the ratios: ${\cal R}_{K/\pi}$, ${\cal R}_{D_s/\pi}$ and ${\cal R}_{\psi(2S)/{J/\psi}}$ in broad agreement with the LHCb data though our predicted ratio ${\cal R}_{\pi /{\mu\nu}}$ is found to be underestimated. The results indicate that the present approach works well in the description of exclusive nonleptonic $B_c$-decays within the framework of the RIQ model.  	
			\end{abstract}
	\maketitle
	\section{Introduction}

	\noindent 
	 The $B_c$-meson is unique because of its two outstanding characteristics features:(1) It is the lowest bound state of two heavy quarks with open(explicit) flavor quantum numbers-a bottom quark(antiquark) and a charm antiquark(quark). The other heavy quark in the standard model i.e, the top quark, can not form a hadron because of its too short life time to be hadronized. (2) It can decay only via weak interactions, since pure strong and electromagnetic interacting processes conserve flavors. Being a ground state of $(b\bar{c})$-system, it lies below the $BD$ meson decay threshold and either of its constituent quarks being heavy which can decay individually, a very rich $B_c$-decay channels are expected with sizable branching ratios and comparatively long life time \cite{A1}. This makes $B_c$-meson an ideal system for studying the heavy quark dynamics.\\
	 
	 Ever since the discovery of $B_c$- meson at Fermilab by the CDF Collaborations \cite{A2} two decades ago, a lot of experimental probes have gone in this sector yielding detection of many ground and excited heavy meson states including the radially excited states : $\psi(2S)$ and $\eta_c(2S)$ \cite{A3}. Many new $B_c$-decay channels have also been observed by the LHCb Collaborations \cite{A4,A5,A6,A7,A8}. The high luminosity available at LHC make it possible to measure various decay channels including those into charmonium states\cite{A9,A10,A11}. Around ${\cal O}(10^9)$ $B_c$-events with a cross section of $1\mu b$ and luminosity of $1fb^{-1}$ \cite{A12} expected at the LHCb are likely to provide sufficient data for a systematic study of the $B_c$ family. Among many observations on $B_c$-decays in recent times, the decay:$B_c\to J/\psi K$ is observed for the first time by the LHCb Collaborations and the measurement of ratios of branching fractions are found to be: 
	 	 \begin{eqnarray}
	 	{\cal R}_{K/\pi}=\frac{{\cal B}(B_c\to J/\psi K)}{{\cal B}(B_c\to J/\psi \pi)}=\begin{cases}
	 	0.069\pm 0.019(Stat.)\pm 0.005(syst.)\ \ \ \ \ \textbf{[13]} \\
 	0.079\pm 0.007(stat.)\pm 0.003(syst.)\ \ \ \ \ \textbf{[14]}	\end{cases}\nonumber 
	 		 \end{eqnarray}
	The ratios of branching fractions: ${\cal R}_{D_s/\pi}$, ${\cal R}_{\psi (2S)/{J/\psi}}$ observed by LHCb Collaborations are:

	\begin{equation}
	{\cal R}_{D_s/\pi}=\frac{{\cal B}(B_c\to J/\psi D_s)}{{\cal B}(B_c\to J/\psi \pi)}=2.9\pm0.57(stat.)\pm0.24(syst.) \ \ \ \ \ \ \textbf{[6]} \nonumber
	\end{equation}
 	 \begin{equation}
 		 	{\cal R}_{\psi(2S)/{J/\psi}}=\frac{{\cal B}(B_c\to \psi(2S)\pi)}{{\cal B}(B_c\to J/\psi \pi)}=0.250\pm 0.068(Stat.)\pm 0.014(Syst.)\pm 0.006\ \ \textbf{[15]}\nonumber
 		 \end{equation}  
 	 \noindent where the last correction term accounts for the uncertainity of $\frac{{\cal B}(\psi (2S)\to  \mu^+\mu^-)}{{\cal B}(J/\psi \to  \mu^+\mu^-)}$. The first measurement of ratio relating the nonleptonic and semileptonic $B_c$-decay rates is also performed by the LHCb Collaborations yielding:
 	 \begin{equation}
 	 	{\cal R}_{\pi^+/{\mu^+\nu}}=\frac{{\cal B}(B_c\to J/\psi\ \pi)}{{\cal B}(B_c\to J/\psi \mu^+\nu_\mu)}=0.049\pm 0.0028(Stat.)\pm 0.0046(Syst.),\ \ \textbf{[16]}\nonumber
 	 \end{equation}
  \noindent which is found at the lower end of the available theoretical predictions.\\
  
  The detection of ground and excited charmonium states and measured observables in the non-leptonic and semileptonic $B_c$-decays to charmonium ground and excited states are of special interest as it is easier to identify them in experiments. The two-body nonleptonic $B_c$-decays have been widely studied using various theoretical approaches and phenomenological models (see the classified bibliography of Ref. \cite{A17}). Most of these studies deals with the $B_c$-meson decay to daughter mesons in their ground states only. Among several theoretical studies on nonleptonic $B_c$-meson decays to radially excited charmonium and charm meson states, the perturbative QCD approach based on $k_{T}$ factorization\cite{A18,A19}, light front quark model using modified harmonic oscillator wave functions\cite{A20}, ISGW2 quark model\cite{A21}, relativistic constituent quark model\cite{A22}, relativistic constituent quark model based on Bethe-Salpeter formalism\cite{A23}, improved instantaneous approximation of the original Bathe-Salpeter equation and mandelstam approach\cite{A24}, perturbative QCD approach\cite{A25}, relativistic instantaneous approximation of the original Bethe-Salpeter equation\cite{A26}, quark model based on improved Bethe-Salpeter approach\cite{A27,A28}, non-relativistic constituent quark model \cite{A29}, relativistic constituent quark model\cite{A30}, relativistic quark model \cite{A31}, QCD factorization using BSW-model and light front quark model \cite{A32}, QCD relativistic quark potential model \cite{A33}, covariant confined quark model\cite{A34} and the relativistic constituent quark model \cite{A35} etc. are noteworthy. In their recent analysis \cite{A19} Zhou Rui $et\ al.$ predicted the ratio between the decay modes: $B_c\to \psi(2S) \pi$ and $B_c\to J/\psi \pi$ in comparison with experimental data within uncertainties and branching fraction of $B_c\to \eta_c(2S)\pi \sim 10^{-3}$ that can hopefully be measured in the LHCb experiment. The recent predictions \cite{A27,A28} of Tian Zhou $et\ al.$ on the branching fractions of radially excited $2S$ and $3S$ charmonium states $\sim 10^{-4}$ lie within the detection accuracy of current experiments. The detection of such decays to radially excited charmonium states, observed ratios of branching fractions and recent predictions of branching fractions in this sector by different model approaches provide us necessary motivation to study these nonleptonic $B_c$-decays within the framework of our relativistic independent quark(RIQ) model. \\
   
   The relativistic independent quark(RIQ) model, developed by our group, has been applied in wide-ranging hadronic sector describing the static properties of hadrons \cite{A36} and their decay properties in the radiative, weak radiative, rare radiative \cite{A37}; leptonic, weak leptonic, radiative leptonic \cite{A38} and semileptonic \cite{A39} decays of mesons. In our recent analysis we predict the magnetic dipole and electromagnetic transitions of $B_c$ and $B_c^*$ mesons in their ground as well as excited states\cite{A40}; the exclusive semileptonic $B_c$-mesons decays to the charmonium ground states in the vanishing \cite{A41} and non-vanishing \cite{A42} lepton mass limit. In this model our group have predicted \cite{A43,A44} the exclusive two body nonleptonic decays of heavy flavored meson to the charmonium and charm mesons in their ground states . We would like to extend the application of the RIQ model to analyze two-body nonleptonic $B_c$-decays to the S-wave charmonium states(nS) along with a light or charm meson state, where $n=1,2,3$ and provide a ready reference to existing and forthcoming experiments. We ignore the decay channels involving higher 4S charmonia, since their properties are stil not understood well.

  The description of nonleptonic decay is notoriously non-trivial as it is strongly influenced by confining color-forces and it involves matrix elements of local four-quark operators in the non-perturbative QCD approach, the mechanism of which is not yet understood well in the Standard model framework. If one ignores the weak annihilation contribution, the nonleptonic transition amplitudes are conveniently described in the so-called naive factorization approximation \cite{A22,A31,A33,A43,A44,A45,A46,A47}, which works reasonably well in two-body nonleptonic $B_c$-decays, where the quark-gluon sea is suppressed in the heavy quarkonium \cite{A32}. Bjorken's intuitive argument on color-transparency in his pioneering work \cite{A48}, theoretical development based on QCD approach in the $\frac{1}{N_c}$ limit \cite{A49} and the heavy quark effective theory (HQET) \cite{A50}, provide justification for such approximation. In the present study we consider the contribution of the current-current operators \cite{A51} only in calculating the tree-level diagram expected to be dominant in these decays. The contribution of the penguin diagram may be significant in the evaluation of CP-violation and search for new Physics beyond Standard model. But its contribution to the decay amplitudes in the present analysis is considered less significant. In fact, the QCD and electroweak penguin operators' contribution has been shown \cite{A52,A53} to be negligible compared to that of current-current operators in these decays due to serious suppression of CKM elements. The Wilson's co-efficients of penguin operators being very small, its contribution to the weak decay amplitude is only relevant in rare decays, where the tree-level contribution is either strongly CKM-suppressed as in $\bar{B}\to \bar{K}^*\pi$ or matrix elements of current-current operators do not contribute at all as in $\bar{B}\to \bar{K}^*\gamma$ and $\bar{B}^0\to \bar{K}^0\phi$ \cite{A51}.\\
      
   The rest of the paper is organized as follows. In section II we introduce the effective Hamiltonian and factorization for two body nonleptonic $B_c $-meson decay modes induced by $b\to c\bar{q}_iq_j$ transition at the quark level. The weak decay form factors representing the hadronic amplitudes, calculated from the overlap integral of meson wave-functions in the framework of the RIQ model are described in Section-III. Section-IV is devoted to the numerical results and discussion and Section-V contains our brief summary and conclusion. A brief review of the RIQ model framework, wave packet representation of the meson state and the momentum probability amplitudes of the constituent quarks inside the meson bound-state are presented in the Appendix.
   
  \section{Effective Hamiltonian and factorization approximation}
  In this section we introduce the effective Hamiltonian for two-body nonleptonic $B_c$-decays induced by $b\to c \bar{q}_iq_j$ at the quark level, where $q_i=u,c$ and $q_j=d,s$. As described above we consider only the contribution of current-current operators at the tree level. Neglecting the contribution of penguin diagrams, the decay modes symbolized by $B_c\to X_{c\bar{c}}(ns)M$, where $X_{c\bar{c}}$ is the ground and excited charmonium states with $n=1,2,3$ and M is a pseudoscalar(P) or a vector (V) meson, are governed by the effective Hamiltonian \cite{A21,A22,A25,A29,A35}:
  \begin{eqnarray}
  	H_{eff}=\frac{G_F}{\sqrt{2}}&&\Biggl\{V_{cb}V_{ud}^{*}\big[c_1(\mu)(\bar{c}b)(\bar{d}u)+c_2(\mu)(\bar{d}b)(\bar{c}u)\big]\nonumber\\
  	&&+V_{cb}V_{cs}^*\big[c_1(\mu)(\bar{c}b)(\bar{s}c)+c_2(\mu)(\bar{s}b)(\bar{c}c)\big]\nonumber\\
  	&&+V_{cb}V_{us}^*\big[c_1(\mu)(\bar{c}b)(\bar{s}b)+c_2(\mu)(\bar{s}b)(\bar{c}u)\big]\nonumber\\
  	&&+V_{cb}V_{cd}^*\big[c_1(\mu)(\bar{c}b)(\bar{d}c)+c_2(\mu)(\bar{d}b)(\bar{c}c)\big]\Biggr\}+h.c\  \ ,
  \end{eqnarray} 
  
  \noindent where ${G_F}$ is the Fermi Coupling constant, $V_{ij}$ are CKM factors; $(\bar{q}_{\alpha}q_{\beta})$ is a short notation for $V-A$ current $q_{\alpha}\gamma^{\mu}(1-\gamma_5)q_{\beta}$, and $c_{1,2}$ are the Wilson coefficients. With the effective Hamiltonian in the form (1), the decay amplitude for $B_c\to X_{c\bar{c}}(nS)M$ is given by 
  \begin{eqnarray}
  	A(B_c\to X_{c\bar{c}}(nS)M)=\langle X_{c\bar{c}}(nS)M|H_{eff}|B_c\rangle
  	=\frac{G_F}{\sqrt{2}}\sum_i \lambda_i C_i(\mu)\langle {\cal O}\rangle_i
  \end{eqnarray}
   where $\lambda_i$ is the CKM factor and $\langle {\cal O}\rangle_i$ is the matrix element of the local four-quark operators. In the framework of naive factorization, the nonleptonic decay amplitude is approximated by the product of two matrix elements of quark currents as:
  
  \begin{equation}
  	\langle X_{c\bar{c}}(nS)M|{\cal O}|B_c\rangle_i=\langle M|J^{\mu}|0\rangle \langle X_{c\bar{c}}(nS)|J_{\mu}|B_c\rangle+(X_{c\bar{c}}(nS)\leftrightarrow M)
  \end{equation}
     \noindent where $J_\mu$ is the weak current. One of these is the matrix element for the $B_c$- transition to one final mesons state, while the other matrix element corresponds to the transition from the vacuum to other final meson state.The latter is given by the corresponding meson decay constant. In this way the hadronic matrix element of four-quark operators can be expressed as the product of decay constant and invariant weak form factors \cite{A21,A29,A43,A44,A47,A54,A55}.\\
     
     Of course there is difficulty inherent in such an approach because the Wilson's coefficients, which include the short-difference QCD effect between $\mu=m_N$ and $\mu=m_b$ are $\mu$-scale and renormalization scheme dependent, while $\langle {\cal O}\rangle_i$ are $\mu$-scale and renormalization scheme independent. As a result, the physical amplitude depends on the $\mu$ scale. However, the naive factorization disentangles the long-distance effects from the short distance sector assuming that the matrix element $\langle {\cal O}\rangle$ at $\mu$ scale, contain non-factorizable contributions in order to cancel the $\mu$ dependence  and scheme dependence of $c_i(\mu)$, i.e., the approximation neglects possible QCD interaction between the meson $M$ and the $B_c X_{c\bar{c}}$ system \cite{A47,A55}. In general, it works in some two-body nonleptonic decays of heavy mesons in the limit of a large number colors. It is expected that the factorization scheme works reasonably well in two-body nonleptonic $B_c$ decays with radially excited charmonium mesons in the final states, where the quark-gluon sea is suppressed in the heavy quarkonium\cite{A21,A32}.\\
     
     We also neglect here the so-called $W$-exchange and annihilation diagram, since in the limit   $M_W\to \infty$, they are connected by Fiertz transformation and are doubly suppressed by the kinematic factor of the order $(\frac{m_i^2}{M_W^2})$. We also discard the color octet current which emerge after the Fiertz transformation of color-singlet operators. Clearly, these currents violate factorization since they can not provide transitions to the vacuum states. Taking into account the Fiertz reordered contribution, the relvant coefficients are not $c_1(\mu)$ and $c_2(\mu)$ but the combination :
     
     \begin{equation}
     	a_{1,2}(\mu)=c_{1,2}(\mu)+\frac{1}{N_c}c_{2,1}(\mu)
     \end{equation}
  
 \noindent Assuming large $N_c$ limit to fix the QCD coefficients $a_1 \approx c_1$ and $a_2\approx c_2$ at $\mu\approx m_b^2$, nonleptonic decays of heavy mesons have been analyzed in Ref.\cite{A21,A25,A33,A56}.\\
  
  The matrix elements corresponding to the transition from vacuum to one of the final state pseudoscalar(P) or vector(V) meson are covariantly expanded in terms of the meson decay constant $f_{P,V}$ as:
  \begin{eqnarray}
  	\langle P|\bar{q}_i^{'}\gamma^\mu \gamma_5 q_j|0\rangle=if_Pp_P^\mu\nonumber\\
  	\langle V|\bar{q}^{'}_i\gamma^\mu q_j|0\rangle=e^{*\mu}f_Vm_V
  \end{eqnarray}
  
  The covariant decomposition of matrix elements of the weak current $J_\mu$ between initial and final pseudoscalar  meson state is 
  
  \begin{eqnarray}
  	\langle P(p_P)|\bar{q}_{c}\gamma_\mu q_{b}|B_c(P)\rangle=&&\big[(p+p_P)_\mu-\frac{M^2-m_P^2}{q^2}q_\mu\big]F_1(q^2)+\frac{M^2-m_P^2}{q^2}q_\mu F_0(q^2)\nonumber\\
  	=&&(p+p_P)_\mu f_+(q^2)+(p-p_P)_\mu f_-(q^2)
  \end{eqnarray}
where 
\begin{eqnarray}
	f_+(q^2)=&&F_1(q^2)\\
	f_-(q^2)=&&\frac{M^2-m^2_P}{q^2} \big[F_0(q^2)-F_1(q^2)\big]
\end{eqnarray}

For transition to the vector meson final state, corresponding matrix element is parametrized as :

\begin{equation}
	\langle V(p_V)|\bar{q}_c\gamma_\mu q_b|B_c(p)\rangle=\frac{2V(q^2)}{M+m_V}i\epsilon_{\mu\nu\rho\sigma}\ e^{*\nu} p^\rho p_V^{\sigma}
\end{equation}
and
\begin{eqnarray}
	\langle V(p_V)|\bar{q}_c\gamma_\mu\gamma_5q_b|B_c(p)\rangle=&&(M+m_V)e^*_\mu A_1(q^2)-\frac{A_2(q^2)}{M+m_V}(e^*.q)(p+p_V)_\mu\nonumber\\
	&&-2m_V\frac{e^*.q}{q^2}q_\mu A_3(q^2)+2m_V\frac{e^*.q}{q^2}q_\mu A_0(q^2)
\end{eqnarray}

where \begin{equation}
	A_3(q^2)=\frac{M+m_V}{2m_V}A_1(q^2)-\frac{M-m_V}{2m_V}A_2(q^2)
\end{equation}

Here $p,p_{P,V}$ stand for the four momentum of the initial and final state meson, respectively. M is the mass of decaying $B_c$ and $m_P$ and $m_V$ stand for the mass of the pseudoscalar and vector mesons, respectively, in the final state. $q=p-p_{P,V}$ denotes the four momentum transfer and $\hat{e}^*$: the polarization of the final state vector meson.\\
In order to cancel the poles at $q^2=0$, invariant weak form factors: $F_0(q^2), F_1(q^2),A_0(q^2)$ and $A_3(q^2)$ satisfy following conditions: 
\begin{equation}
	F_0(0)=F_1(0)\ \ \ \ and\ \ \ A_0(0)=A_3(0)\nonumber
\end{equation} 

The decay rate for nonleptonic transition $M\to P_1P_2$ is expressed in terms of the decay amplitude $A(B_c\to P_1P_2)$ as:
\begin{equation}
	\Gamma(B_c\to P_1P_2)=\frac{|\vec{k}|}{8\pi M^2}|A(B_c\to P_1P_2)|^2
\end{equation}
where $\vec{k}$ is the magnitude of three-momentum of the final state meson. In the parent meson rest frame it is given by 
\begin{equation}
	|\vec{k}|=|\vec{p}_{P_1}|=|\vec{p}_{P_2}|=\frac{1}{2M}\biggl\{\big[M^2-(m_{P_1}+m_{P_2})^2\big]\big[M^2-(m_{P_1}-m_{P_2})^2\big]\biggr\}^{1/2}
\end{equation}
The corresponding expression of the decay rate for $M\to PV(VP)$ is obtained in the form:
\begin{equation}
	\Gamma(B_c\to PV,VP)=\frac{|\vec{k}|^3}{8\pi m_V^2}|A(B_c\to PV,VP)|^2
\end{equation}
with 
\begin{equation}
	|\vec{k}|=\frac{1}{2M}\biggl\{\big[M^2-(m_{P,V}+m_{V,P})^2\big]\big[M^2-(m_{P,V}-m_{V,P})^2\big]\biggr\}^{1/2}
\end{equation}
The relevant decay amplitude (2) is then expressed in the form:
\begin{eqnarray}
	A=&&\frac{G_F}{\sqrt{2}}(CKM factor)(QCD factor)(M^2-m_{P_1}^2)\times f_{P_2}F_0^{B_c\to P_1}(q^2),\nonumber\\
	A=&&\frac{G_F}{\sqrt{2}}(CKM factor)(QCD factor)2m_Vf_VF_1^{B_c\to P}(q^2)\nonumber\\
	and\nonumber\\
	A=&&\frac{G_F}{\sqrt{2}}(CKM factor)(QCD factor)2m_Vf_PA_0^{B_c\to V}(q^2)\nonumber
\end{eqnarray}

\noindent for $B_c\to P_1P_2$, $B_c\to PV$ and $M\to VP$ decay, respectively. \\
The factorized amplitudes(3) expressed in terms of meson decay constants $(f_{P,V})$ and weak form factors $F_0^{B_c\to P_1},\ F_1^{B_c\to P},\ A_0^{B_c\to V}$ it is straight forward to predict the decay rate for different decay processes in the RIQ model framework.\\

\section{transition amplitude and weak decay form factors in the relativistic independent quark model.}
 We study two-body nonleptonic $B_c$-decays in three categories: $B_c\to PP$, $B_c\to PV$ and $B_c\to VP$, where P and V stand for pseudoscalar and vector meson final states, respectively. The decay amplitude is calculated here from relevant tree-level diagram as shown in Fig.1. The color-favored "class-I" decays, represented by Fig 1.a, are characterized by external $W$-emission, where the decay amplitude is proportional to the QCD-factor $a_1(\mu)$. However the "class-III" type decays, represented in Fig 1.c, are those in which both the QCD factors: $a_1(\mu)$ and $a_2(\mu)$ interfere, providing effective contribution to the factorized decay amplitude. As described above we consider the two-body nonleptonic $B_c$-decays induced by $b\to c\bar{q}_iq_j$ transition at quark level, where $\bar{q}_i=\bar{u},\bar{c}$ and $q_j=d,s$, with $\bar{c}$-antiquark remaining a spectator. In the present study we restrict our discussion to class I and class III $ B_c$-decay modes: each mode involving either $\eta_c(nS)$ or $\psi(nS)$in the final state.\\
 
In fact the decay process physically occurs in the momentum eigen-states of participating mesons. Therefore, in a field-theoretic description of the decay process, it is appropriate to represent the meson bound-state in terms of a momentum wave-packet reflecting momentum and spin distribution between constituent quark and antiquark in the meson-core. In RIQ model, the wave packet corresponding to a meson bound-state $|B_c(\vec{p},S_{B_c})$, for example, at a definite moment $\vec{p}$ and spin state $S_{B_c}$ is represented by:
 \begin{eqnarray}
 	|B_c(\vec{p},S_{B_c})\rangle =&&\hat{\Lambda}(\vec{p},S_{B_c})|(\vec{p}_b,\lambda_b);(\vec{p}_{\bar{c}},\lambda_{\bar{c}})\rangle\\
 	=&&\hat{b}^{\dagger}_b(\vec{p}_b,\lambda_b)\tilde{b}_c^{\dagger}(\vec{p}_c,\lambda_c)|0\rangle \nonumber
 \end{eqnarray}
where $|(\vec{p}_b,\lambda_b);(\vec{p}_{\bar{c}},\lambda_{\bar{c}})\rangle$ is the Fock-space representation of the unbound quark and antiquark in a color-singlet configuration with respective momentum and spin : $(\vec{p}_b,\lambda_b)$ and $(\vec{p}_c\lambda_c)$. $\hat{b}^{\dagger}_q(p_b,\lambda_b)$ and $\hat{\tilde{b}}_c(\vec{p}_c,\lambda_c) $ are, the quark and antiquark creation operator. Here $\hat{\Lambda}(\vec{p},S_{B_c})$ is a bag like integral operator taken in the form:

\begin{equation}
	\hat{\Lambda}(\vec{p},S_{B_c})=\frac{\sqrt{3}}{\sqrt{N(\vec{p})}}\sum_{\lambda_b,\lambda_c}\zeta^{B_c}_{b,c}(\lambda_b,\lambda_c)\int d^3p_bd^3p_c \delta^{(3)}(\vec{p}_b+\vec{p}_c-\vec{p}){\cal G}_{B_c}(\vec{p}_b,\vec{p}_c)
\end{equation}
Here $\sqrt{3}$ is the effective color factor and $\zeta^{B_c}_{b,c}(\lambda_b,\lambda_c)$ is the SU(6) spin-flavor coefficients for $B_c$-meson state. Imposing the normalization condition in the form $\langle B_c(\vec{p})|B_c(\vec{p}^{'}) \rangle =\delta^{(3)}(\vec{p}-\vec{p}^{'})$, the meson state normalization $N(\vec{p})$ is obtainable in an integral form:
\begin{equation}
	N(\vec{p})=\int d^3\vec{p}_b |{\cal G}_{B_c}(\vec{p}_b,\vec{p}-\vec{p}_b)|^2
\end{equation}
Finally ${\cal G}_{B_c}(\vec{p}_b,\vec{p}-\vec{p}_b)$ denote the momentum distribution function for the quark and antiquark pair in the meson core. In this model ${\cal G}_{B_c}(\vec{p}_b,\vec{p}-\vec{p}_b)$ is taken in the form:${\cal G}_{B_c}(\vec{p}_b,\vec{p}-\vec{p}_b)=\sqrt{G_b(\vec{p}_b)G_c(\vec{p}-\vec{p}_b)}$ in the straight forward extension of the ansatz of Margolis and Mendel in their bag model description \cite{A57}; where $G_b(\vec{p}_b)$ and $G_c(\vec{p}-\vec{p}_b)$ refer to individual momentum probability amplitude of the constituent quark. Here the effective momentum distribution function in fact embodies the bound-state character in $|B_c(\vec{p},S_{B_c})\rangle$.

 \begin{figure}[!hbt]
	\includegraphics[width=1\textwidth]{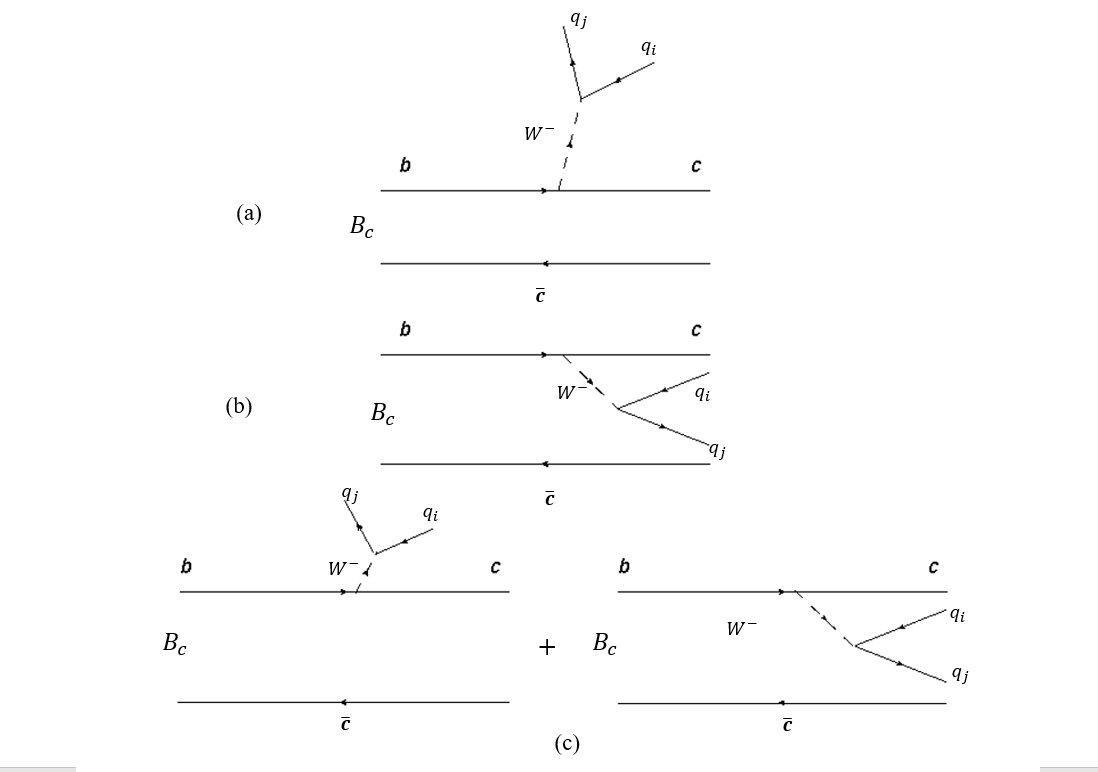}
	\caption{Quark level diagram of nonleptonic decay of meson $B_c\to X_{c\bar{c}}(nS)M$}
\end{figure}

Any residual internal dynamics responsible for the decay process can, therefore, be described at the constituent level by the otherwise unbound quark and antiquark using usual Feynman technique. The constituent level $S$-matrix element $S_{fi}^b\to c\bar{q}_iq_j$ obtained from the appropriate Feynman diagram, when operated upon by the bag like operator $\hat{\Lambda}(\vec{p},S_{B_c})$ (17) can give rise to the mesonic level $S$-matrix  in the form:

\begin{equation}
	S_{fi}^{B_c\to X_{c\bar{c}}(nS)M}\to \hat{\Lambda}(\vec{p},S_{B_c})S_{fi}^{b\to c \bar{q}_iq_j}
\end{equation}

\subsection{$B_c\to P_1P_2$}
The nonleptonic decay mode: $B_c\to P_1P_2$, where $P_1$ and $P_2$ are pseudoscalar mesons, [Fig 1.(a)], is induced by $b\to c\bar{q}_iq_j$ transition at quark level with emission of $W$-boson. The resulting quark $c$ and the spectator anti-quark $\bar{c}$ attached to the decaying meson state $|B_c(\vec{p},S_{B_c})\rangle$ hadronize to form $|P_1(\vec{k},S_{P_1})\rangle$, while the externally emitted $W$-boson with four momentum $q$, which decays to a quark-antiquark pair, subsequently hadronize to form the pseudoscalar meson state : $|P_2(\vec{q},S_{P_2})\rangle$. Considering the wave packet representation of the participating meson states in the factorized decay amplitude (3), the  $S$-matrix element for $B_c\to P_1P_2$ can be obtained in the general form:
   
\begin{equation}
	S_{fi}=(2\pi)^4 \delta^{(4)}(p-q-k)(-i{\cal M}_{fi})\times\frac{1}{\sqrt{V2E_{B_c}}}\prod_f \frac{1}{V2E_f}
\end{equation}  

\noindent The invariant transition amplitude ${\cal M}_{fi}$ is infact extracted in the form :

\begin{equation}
	{\cal M}_{fi}=\frac{G_F}{\sqrt{2}}V_{bc}V_{\bar{q}_iq_j}a_1 A
\end{equation}
where, $A=h^{\mu}H_{\mu}$with
\noindent\begin{equation}
	h^{\mu}=\sqrt{\frac{2E_{P_2}}{(2\pi)^3}}\langle P_2(\vec{q},S_{P_2})|J^{\mu}|0\rangle
\end{equation}
and 
\begin{eqnarray}
	H_{\mu}=\frac{1}{\sqrt{N_{B_c}(\vec{p})N_{P_1}(\vec{k})}}\int{\frac{d^3{\vec{p}_b}G_{B_c}(\vec{p}_b,\vec{p}-\vec{p}_b)G_{P_1}(\vec{p}_b+\vec{k}-\vec{p},\vec{p}-\vec{p}_b)}{\sqrt{E_b(\vec{p}_b)E_c(\vec{p}_b+\vec{k}-\vec{p})}}}\nonumber\\
	\times\sqrt{\big[E_b(\vec{p}_b)+E_c(\vec{p}-\vec{p}_b)\big]\big[E_c(\vec{p}_b+\vec{k}-\vec{p})+E_{\bar{c}}(\vec{p}-\vec{p}_b)\big]}\langle S_{P_1}|J_{\mu}(0)|S_{B_c}\rangle
\end{eqnarray}
Here $E_b(\vec{p}_b)$ and $E_c(\vec{p}_b+\vec{k}-\vec{p})$ stand for energy of the non-spectator quark of the parent and daughter meson; $\vec{p}_b,\vec{p},\vec{k}$ represent three momentum of the non-spectator constituent quark $b$, parent meson $B_c$ and the daughter meson $P_1$, respectively. $q=p-k$ is equal to the four momentum associated with the meson state $|P_2(\vec{q},S_{P_2})\rangle$. $\langle S_{P_1}|J_{\mu}|S_{B_c}\rangle$ represents symbolically the spin matrix elements of the effective vector-axial vector current. For transitions involving non-spectator constituent quark $b$, the spin matrix element is: 
\begin{equation}
	\langle S_{P_1}|J_{\mu}|S_{B_c}\rangle=\sum_{\lambda_b,\lambda^{'}_{c}\lambda_{\bar{c}}}\zeta_{b,\bar{c}}^{B_c}(\lambda_{b},\lambda_{\bar{c}})\zeta_{c,\bar{c}}^{P_1}(\lambda_{c}^{'},\lambda_{\bar{c}}) \bar{u}_c(\vec{p}_{b}+\vec{k}-\vec{p},\lambda_{c}^{'})\gamma_\mu(1-\gamma_5)u_b(\vec{p}_b,\lambda_b)
\end{equation}
 Here $u_i$ stands for free Dirac spinor. $\zeta^{B_c}(\lambda_{b},\lambda_{\bar{c}})$ and $\zeta^{P_1}(\lambda_{c}^{'},\lambda_{\bar{c}})$ are the appropriate SU(6) spin-flavor coefficients corresponding to the parent and daughter meson, respectively. \\
 
 It may be pointed out here that in the description of the decay processes: $B_c\to P_1P_2$ in the RIQ model framework, the three-momentum conservation is ensured explicitly via $\delta^{(3)}(\vec{p}_q+\vec{p}_{\bar{q_2}}-\vec{p})$ in the participating meson states. However energy conservation in such a scheme is not ensured so explicitly. This is in fact a typical problem in all potential model descriptions of meson as a bound-state of valence quark and antiquark interacting via some instantaneous potential. This problem has been addressed in our previous analysis in the context of radiative leptonic decays of heavy flavored means: $B,B_c,D,D_s$ \cite{A38}, where the effective momentum distribution function ${\cal G}_M(\vec{p}_{q_1},\vec{p}_{\bar{q}_2})$ that embodies the bound-state characteristics of the meson, ensures energy conservation in an average sense satisfying $E_M=\langle M(\vec{p},S_M)|\big[E_{q_1}(\vec{p}_{q_1})+E_{\bar{q}_2}(\vec{p}_{\bar{q}_2})\big]|M(\vec{p},S_M)\rangle$. In view of this we take the energy conservation constraint: $M=E_{q_1}(\vec{p}_{q_1})+E_{\bar{q}_2}(-\vec{p}_{\bar{q}_1})$ denoting the mass of the meson at rest. This along with the 3-momentum conservation via appropriate $\delta^{(3)}(\vec{p}_{\bar{q}_1}+\vec{p}_{q_2}-\vec{p})$ in the meson state ensure required 4-momentum conservation:$\delta^{(4)}(p-k-q)$ at the mesonic level, which is pulled out of the quark level integration so as to obtain the $S$-matrix element in the standard form (20).\\

 Since the axial vector current does not contribute to the decay amplitude in the decay processes: $B_c\to P_1P_2$, the only non-vanishing vector current part of (23) is simplified after calculating corresponding spin matrix elements (24) using usual spin algebra. The resulting time-like and space-like part of the hadronic matrix element in the parent meson rest-frame are obtained, respectively, as:
 \begin{equation}
 	\langle P_1(\vec{k})|V_0|B_c(0)\rangle =H_0=\int d\vec{p}_bC(\vec{p}_b)\bigl\{\big[E_b(\vec{p}_b)+m_b\big]\big[E_{\bar{c}}(\vec{p}_b+\vec{k})+m_{\bar{c}}\big]+\vec{p}_b^2\bigr\}
 \end{equation}
and 
\begin{equation}
	\langle P_1(\vec{k})|V_i|B_c(0)\rangle =H_i=\int d\vec{p}_b\ C(\vec{p}_b)\ \big[E_b(\vec{p}_b)+m_b\big]k_i	\ \ ,
\end{equation}
  
  \noindent where 
  \begin{equation}
  	C(\vec{p}_b)=\frac{{\cal G}_{B_c}(\vec{p}_b,-\vec{p}_b){\cal G}_{P_1}(\vec{p}_b+\vec{k},-\vec{p}_b)}{\sqrt{N_{B_c}(0)N_{P_1}(\vec{k})}}\sqrt{\frac{\big[E_b(\vec{p}_b)+E_{\bar{c}}(-\vec{p}_b)\big]\big[E_c(\vec{p}_b+\vec{k})+E_{\bar{c}}(-\vec{p}_b)\big]}{E_b(\vec{p}_b)E_c(\vec{p}_b+\vec{k})\big[E_b(\vec{p}_b)+m_b\big]\big[E_c(\vec{p}_b+\vec{k})+m_c\big]}}
  \end{equation}

\noindent Now a comparison of the results (25-27) with the corresponding expression of the covariant factorized amplitude (6-8) yields the Lorentz invariant form factors $f_{\pm}(q^2)$ in the form:
\begin{eqnarray}
	f_{\pm}(q^2)=\frac{1}{2}\int d\vec{p}_b C(\vec{p}_b)&&\biggl\{\big[E_b(\vec{p}_b)+m_b\big]\big[E_c(\vec{p}_b+\vec{k})+m_c\big]+\vec{p}_b^2\nonumber\\
	&&\pm\big[E_b(\vec{p}_b)+m_b\big]\big[M\mp E_{P_1}\big]\biggr\}
\end{eqnarray}

  \noindent Then it is straightforward to get the model expression for the form factor $F_0(q^2)$ from $F_0(q^2)=\bigg[\frac{q^2}{(M^2-m_{P_1}^2)}\bigg]f_-(q^2)+f_+(q^2)$ in terms of which the decay rate $\Gamma(B_c\to P_1P_2)$ is obtained as 
  \begin{equation}
  	\Gamma(B_c\to P_1P_2)=\frac{|\vec{k}|}{8\pi M^2}|A_1|^2|F_0(q^2)|^2
  \end{equation}
where

\begin{equation}
	|A_1|=\frac{G_F}{\sqrt{2}}V_{bc}V_{\bar{q}_iq_j}a_1(M^2-m_{P_1}^2)f_{P_2}
\end{equation}  

\subsection{$B_c\to PV$}
In the nonleptonic decay process $B_c\to PV$ of class I category, the externally emitted $W$-boson first decays to a quark-antiquark pair which ultimately hadronize to a vector meson(V) and the $c$ quark originating from the nonspectator $b$ decay along with the spectator $\bar{c}$, hadronize to the pseudoscalar meson (P) forming a member of the charmonium family. It can be readily checked that the decay amplitude in the $B_c$-rest frame in such decays is obtainable in terms of the invariant form factor $F_1(q^2)$ as:

\begin{equation}
	\langle P(\vec{k})V(\vec{q})|{\cal H}_{eff}|B_c(0)\rangle = i\frac{G_F}{\sqrt{2}}V_{bc}V_{\bar{q}_iq_j}2a_1m_Vf_VF_1(q^2)(e^*.p)
\end{equation}  

\noindent Here $e^*$ denotes the polarization vector associated with the daughter meson(V). The form factor $F_1(q^2)$ in the parent meson rest frame is obtained as:

\begin{equation}
	F_1(q^2)=f_+(q^2)=\frac{1}{2} \int d\vec{p}_b C(\vec{p}_b)\biggl\{\big[E_b(\vec{p}_b)+m_b\big]\big[E_c(\vec{p}_b+\vec{k})+m_c\big]+\vec{p}_b^2+\big[E_b(\vec{p}_b)+m_b\big]\big[M-E_P\big]\biggr\}
\end{equation}
in terms of which the decay rate $\Gamma(B_c\to PV)$ is expressed as 
\begin{equation}
	\Gamma(B_c\to PV)=\frac{|\vec{k}|^3}{8\pi m_V^2}|A_2|^2|F_1(q^2)|^2
\end{equation}
where

\begin{equation}
	|A_2|=\frac{G_F}{\sqrt{2}}V_{bc}V_{\bar{q}_iq_j}2a_1m_Vf_V
\end{equation}

\subsection{$B_c\to VP$}

In the decay process of this category, the externally emitted $W$-boson first decays to a quark-antiquark pair which ultimately hadronize to a pseudoscalar meson(P). The resulting $c$ from the non-spectator $b$-decay and the spectator $\bar{c}$ hadronize to the vector meson(V) belonging to the charmonium family. In this case the vector current does not contribute and the non-vanishing decay amplitude due to axial-vector current in $B_c$-rest frame is obtained in a simple form:

\begin{equation}
	\langle V(\vec{k})P(\vec{q})|{\cal H}_{eff}|B_c(0)\rangle =i\frac{G_F}{\sqrt{2}}V_{bc}V_{\bar{q}_iq_j}2a_1m_Vf_PA_0(q^2)(e^*.p)
\end{equation}

\noindent Although all four invariant form factors: $A_1, A_2,A_3$ and $A_0$ are expected to contribute to the decay amplitude in these decays , the contribution of a single form factor $A_0(q^2)$ is relevant here as shown in (35). This is due to the mutual cancellation of terms arising from the linear relation(11). With the appropriate wave packet representation of the participating meson states: $|V(\vec{k},S_V)\rangle$ and $|B_c(0,S_{B_c})\rangle$, the non-vanishing factorized amplitude $\langle V(\vec{k},S_V)|A_{\mu}|B_c(0,S_{B_c})\rangle$ is calculated with respect to three spin states $(S_V=\pm1,0)$ of the vector meson(V) in the final state. In the calculation of the spin matrix element, the polarization vector $e^*$ associated with the final state vector meson is extracted from the model dynamics. The model expressions of the time-like and space-like part of the decay amplitude are then obtained in the parent meson rest frame as:

\begin{equation}
	\langle V(\vec{k},S_V)|A_0|B_c(0,S_{B_c})\rangle=\int d\vec{p}_b\ C(\vec{p}_b)\big[E_b(\vec{p}_b)+m_b\big](\hat{e}^*.\vec{k})
\end{equation} 
and 
\begin{equation}
	\langle V(\vec{k},S_V)|A_i|B_c(0,S_{B_c})\rangle =\int d\vec{p}_b\ C(\vec{p}_b)\biggl\{\big[E_b(\vec{p}_b)+m_b\big]\big[E_c(\vec{p}_b+\vec{k})+m_c\big]-\frac{\vec{p}_b^2}{3}\biggr\}\hat{e}^* \ ,
\end{equation}

\noindent respectively. A comparison of the expressions in (36,37) with the covariant expansion (9,10) leads to the model expression of the relevant form factor $A_0(q^2)$ as:

\begin{equation}
	A_0(q^2)=\frac{1}{2}\int d\vec{p}_b\ C(\vec{p}_b)\biggl\{\big[E_b(\vec{p}_b)+m_b\big]\big[M-E_V\big]\big[E_b(\vec{p}_b)+m_b\big]\big[E_c(\vec{p}_b+\vec{k})+m_c\big]-\frac{\vec{p}_b^2}{3}\biggr\}
\end{equation}

\noindent Then the decay rate $\Gamma(B_c\to VP)$ in terms of $A_0(q^2)$ is obtained in a straight forward manner as:
\begin{equation}
	\Gamma(B_c\to VP)=\frac{|\vec{k}|^3}{8\pi m_V^2}|A_3|^2|A_0(q^2)|^2
\end{equation}

\noindent where

\begin{equation}
	|A_3|=\frac{G_F}{\sqrt{2}}V_{bc}V_{\bar{q}_iq_j}2a_1m_Vf_P
\end{equation}

\noindent The two body nonleptonic $B_c$-decay described so far in this section refer to the color-favored "class-I" decays involving external emission of $W$-boson [Fig. 1(a)]. For class III decay modes considered in the present study, the contribution to the decay amplitude is extracted from the Pauli interence of both the diagrams depicting external and internal emission of $W$-boson. The model expressions for relevant form factors and decay rates for such decays can be obtained by suitably replacing the relevant flavor degree of freedom, quark masses, quark binding energies, QCD factor $a_1$, $a_2$ and the decay constants.

\section{Numerical results and discussion}

For calculating the two-body nonleptonic $B_c$-decays in the relativistic independent quark(RIQ) model, we need to fix the flavor-independent potential parameters $(a,V_0)$, quark masses$(m_q)$ and corresponding quark binding energy $(E_q)$. In fact, these parameters have already been fixed in our model in reproducing the experimental meson spectra in the light and heavy flavor sector\cite{A36} and subsequently used in the description of a wide ranging hadronic phenomena\cite{A37,A38,A39,A40,A41,A42,A43,A44} involving participating mesons in their ground state. Accordingly the potential parameters used in the present study are:
\begin{equation}
(a,V_0)=(0.17166\ GeV^3,-0.1375\ GeV)
\end{equation}

\noindent The quark masses and corresponding binding energies in $GeV$ are taken as: 
\begin{eqnarray}
m_u=m_d=&&0.07875 \ \ \ \ \ \ \ E_u=E_d=0.47125\nonumber\\ 
m_s=&&0.31575 \ \ \ \ \ \ \ \ \ \ \ \ \ \ \ E_s=0.591\nonumber\\
m_c=&&1.49275 \ \ \ \ \ \ \ \ \ \ \ \ \ \ \ E_c=1.57951\nonumber\\
m_b=&&4.77659 \ \ \ \ \ \ \ \ \ \ \ \ \ \ \ E_b=4.76633
\end{eqnarray}

\noindent For relevant CKM parameters and lifetime of $B_c$-meson, we take their central values from PDG \cite{A58} as:
\begin{eqnarray}
|V_{cb}|=&&0.041,\ \ \ \ |V_{ud}|=0.9737,\ \ \ \ |V_{cs}|=0.987\nonumber\\
|V_{us}|=&&0.2245,\ \ \ \ |V_{cd}|=0.221;\ \ \ \ \tau(B_c)=0.51\ ps 
\end{eqnarray}
For the masses and decay constants of the participating mesons, considered as phenomenological inputs in the present calculation, we take their central values of the available observed data from Ref.\cite{A58,A59,A60}. In the absence of the observed data in the charmonium and charm meson sector, we take the corresponding predicted values from established theoretical approaches\cite{A61,A62,A63,A64}. Accordingly the updated meson masses and decay constants used in the present analysis are listed in Table I.

\begin{table}[!hbt]
	\renewcommand{\arraystretch}{1}
	\centering
	\setlength\tabcolsep{5pt}
	\caption{The masses and decay constants of mesons}
	\label{tab1}
	\begin{tabular}{ccccc}
		\hline
		\hline Particle& &Mass\cite{A58} &Decay constant&Reference  \\
		&&(MeV)&(MeV)&\\
		\hline$\pi$&&139.57&130.5& \cite{A58}\\
		$\rho$&&775.11&221&\cite{A58}\\
		$K^\pm$&&493.677&155.72& \cite{A58}\\
		$K^{\pm*}$&&891.67&220& \cite{A58}\\
		$D^\pm(1S)$&&1869.5&205.8& \cite{A58}\\
		$D^*(1S)$& &2010.2&252.2&\cite{A63} \\
		$D_s^\pm(1S)$& &1968.35&252.4& \cite{A58} \\
		$D_s^{*\pm }(1S)$&  &2112.2&305.5&\cite{A63} \\
		$\eta_c (1S)$& & 2983.9&387&\cite{A64} \\
		$J/\psi (1S)$& & 3096.9 &418&\cite{A64} \\
		$B_c$& &6274.47& &\\
		\hline
		\hline Particle && Mass (MeV)&Reference&\\	
		\hline
		
		$D(2S)$&&2581&\cite{A61}&\\
		$D^{*\pm}(2S)$&&2637&\cite{A59}&\\
		$D_s(2S)$ &&2673&\cite{A61}&\\
		$D_s^*(2S)$&&2732&\cite{A60}&\\
		$\eta_c(2S)$&&3637.&\cite{A58}&\\
		$\psi(2S)$&&3686.1&\cite{A58}&\\				
		$D^\pm(3S)$	&&3068&\cite{A61}&\\
		$D^{\pm*}(3S)$&&3110&\cite{A61}&\\
		$D_s^{\pm }(3S)$&&3154&\cite{A61}&\\
		$D_s^{\pm *}(3S)$&&3193&\cite{A61}&\\
		$\eta_c(3S)$&&4.007&\cite{A62}&\\
		$\psi(3S)$&&4039.1&\cite{A58}&\\
				\hline
	\end{tabular}
\end{table}

Note that, in the prediction of nonleptonic decays, uncertainties creep into the calculation through input parameters: model parameters, CKM parameters, meson decay constants and QCD co-efficients$(a_1,a_2)$ etc. As mentioned above we use in our calculation, the potential parameters(41) and the quark masses and quark binding energies(42)  that have already been fixed at the static level application of the RIQ model by fitting the mass spectra of mesons in their ground state\cite{A36}. The same set of parameters have been used in the earlier application of RIQ model yielding adequate description of a wide ranging hadronic phenomena in the light and heavy flavor mesons in their ground state. Subsequently we extend our model application to study the $B_c$-meson decays into radially excited daughter mesons \cite{A40}. In the calculation of such decay processes \cite{A40} we use the same set of input parameters(41,42) except the quark binding energies. The quark binding energies for excited meson states, are obtained by solving the equation representing appropriate binding energy condition in our model. As such we do not use any free parameters that could be fine-tuned from time to time to predict wide ranging hadronic phenomena as stated above. In that sense, we perform parameter-free calculations in our studies. In order to avoid uncertainties that might creep into our calculation through the CKM parameters and decay constants, we take their central values of the observed data from Ref \cite{A58}. In those cases where observed data for decay constants are not available, we use the predicted data from established model and theoretical approaches\cite{A63,A64}.\\
\noindent As regards QCD coefficient$(a_1,a_2)$, different values have been used in the literature, in the calculation of the nonleptonic transitions of $B_c$-mesons induced by $b$-quark decay. For example Colangelo $\it{et al.}$, in Ref \cite{A25} use QCD co-efficients Set(1): $(a_1^b,a_2^b)=(1.12,-0.26)$ as fixed in Ref \cite{A65}, whereas in most of earlier calculations, authors use a different set of QCD co-efficients Set 2:$(a_1^b,a_2^b)=(1.14,-0.2)$ fixed by Buras ${\it et\ al.}$ in mid 1980s. We use both the sets of QCD coefficients in our calculation.\\

Before calculating invariant form factors using our input parameters(41,42), we would like to elaborate a bit on the energy conservation ansatz used here to ensure the required energy-momentum conservation in the description of nonleptonic $B_c$-meson decays. Considering a meson state $|X(0)\rangle $ decaying at rest, the energy conservation constraint: $M=E_{q_1}(\vec{p}_{q_1})+E_{q_2}(-\vec{p}_{q_1})$ might lead to spurious kinematic singularities at the quark level integration that appear in the decay amplitude. This problem has been addressed in our model analysis on radiative leptonic decays of heavy flavored mesons \cite{A38} and in similar studies based on QCD relativistic quark model approach \cite{A25}, by assigning a running mass to the non-spectator quark $q_1$:
\begin{equation}
m_{q_1}^2(|\vec{p}_{q_1}|)=M^2-m_{q_2}^2-2M\sqrt{|\vec{p}_{q_1}|^2+m_{q_2}^2}\nonumber
\end{equation} 
as an outcome of the energy conservation ansatz, while retaining definite mass $m_{q_2}$ of the spectator quark $\bar{q}_{2}$. This leads to an upper bound on the quark momentum $|\vec{p}_{q_1}|<\frac{M^2-m_{q_2}^2}{2M} $ in order to retain $m_{q_1}^2(|\vec{p}_{q_1}|)$ positive definite. The shape of radial momentum distribution amplitude: $|\vec{p}_{q_1}|{\cal G}_X(\vec{p}_{q_1},-\vec{p}_{q_1})$ over allowed kinematic range: $0\le |\vec{p}_{q_1}|<|\vec{p}_{q_1}|_{max}$ as shown in Fig.(5-7) also match with the shape obtained in similar studies in the QCD relativistic quark model approach \cite{A25}. The rms value of the active quark momentum: $\sqrt{\langle |\vec{p}_{q_1}^2|\rangle }$, where $\langle |\vec{p}_{q_1}^2|\rangle = \langle X(0)|\vec{p}_{q_1}^2|X(0)\rangle$; the expectation value of the binding energies of the active quark $q_1$, and spectator $q_2$ and sum of the binding energy of quark and antiquark pair: $\langle E_{q_1}(\vec{p}_{q_1}^2)\rangle$, $\langle E_{q_2}(|-\vec{p}_{q_1}|^2)\rangle$ and  $\langle E_{q_1}(\vec{p}_{q_1}^2)+E_{q_2}(|-\vec{p}_{q_1}|^2)\rangle$, respectively, calculated in the framework of RIQ model, are presented in Table II.

\begin{table}[!hbt]
	\renewcommand{\arraystretch}{1}
	\centering
	\setlength\tabcolsep{5pt}
	\caption{The rms values of quark momentum, expectation values of the quark and antiquark and expectation value of sum of the quark and antiquark pair in the meson states :}
	\label{tab2}
	\begin{tabular}{cccccc}
		\hline
		\hline Meson state &$\sqrt{\langle \vec{p}_{q_1}^2}\rangle$ &$\langle E_{q_1} (\vec{p}_{q_1}^2) \rangle$ &$\langle E_{q_2} (|-\vec{p}_{q_1}^2|) \rangle$& $\langle [E_{q_1} (\vec{p}_{q_1}^2)+E_{q_2} (|-\vec{p}_{q_1}^2|) \rangle$& Observed meson  \\
		$|X(0)\rangle$ &(GeV)&(GeV)&(GeV)&(GeV)&mass(GeV)\\
		\hline
	$|B_{u}(0)\rangle$& 0.51 & 4.799& 0.480& 5.279& 5.27925\\
		$|B_c(0)\rangle$ & 0.66 & 4.657 & 1.629& 6.286& 6.27447\\
			$|D(0)\rangle$ & 0.4506& 1.4418& 0.4275 &1.8693& 1.86965\\
				$|D_s(0)\rangle$ & 0.4736&1.4165&0.5517&1.9682&1.96835\\\hline
				\hline

		\hline
	\end{tabular}
\end{table}

It is note worthy to discuss three important aspects of our results in Table II.\\
(1) The rms value of the quark momentum in the meson bound states is much less than the corresponding upper bound $|\vec{p}_{q_{1}}|_{max}$ as expected. (2) The average energy of constituent quark of same flavor in different meson bound states do not exactly match. This is because the kinematics and binding energy condition for constituent quarks due to color forces involved are different from one meson bound state to other. The constituent quarks in the meson bound state are considered to be free particles of definite momenta, each associated with its momentum probability amplitude derivable in this model via momentum space projection of the respective quark eigen-modes. On the other hand the energy shown in Eq(42), which are energy-eigen values of the corresponding bound quarks with no definite momenta of their own, are obtained in RIQ model from respective quark orbitals by solving the Dirac equation. No wonder to find the marginal difference between the energy eigen values (42) and average energy of constituent quarks shown in Table II. (3) Finally we obtained the expectation values of the sum of the energy of constituent quark and antiquark in the meson bound state in good agreement with corresponding observed masses as shown in Table II. These important aspects of our results lend credence to our energy conservation ansatz in an average sense through the effective momentum distribution function like ${\cal G}_X(\vec{p}_{q_1},-\vec{p}_{q_1})$ in the meson bound state $|X(0)\rangle $. This ansatz along with 3 momentum conservation in the meson bound state ensures the required energy momentum conservation in our description of several decay processes \cite{A36,A37,A38,A39,A40,A41,A42,A43,A44} as pointed out earlier. In the absence of any rigorous field theoretic description of the meson bound states, invoking such an ansatz is no doubt a reasonable approximation, for constituent level description of hadronic phenomena.

\begin{figure}[!hbt]
	\includegraphics[width=.3\textwidth]{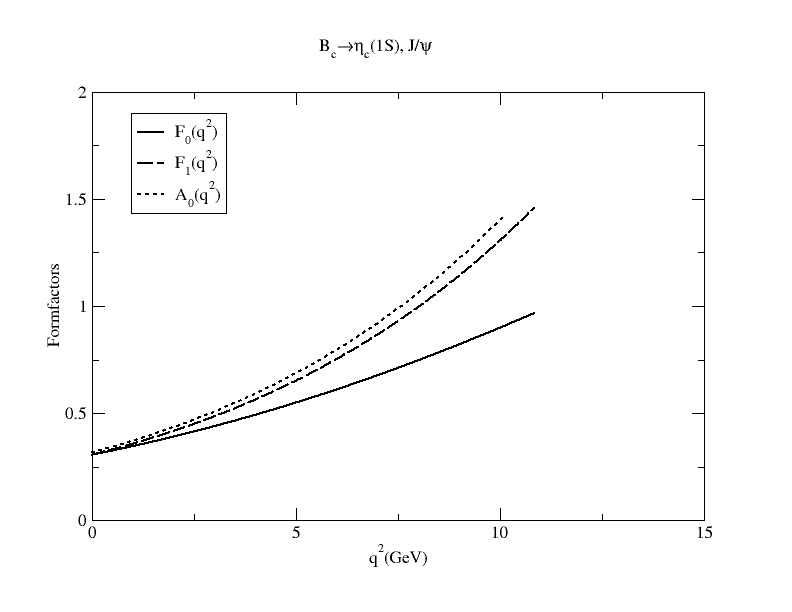}
	\includegraphics[width=.3\textwidth]{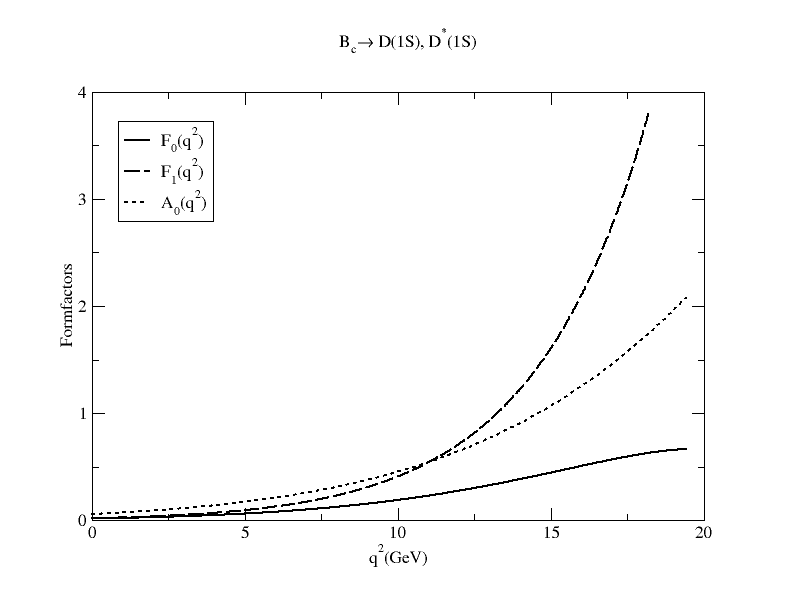}
	\includegraphics[width=.3\textwidth]{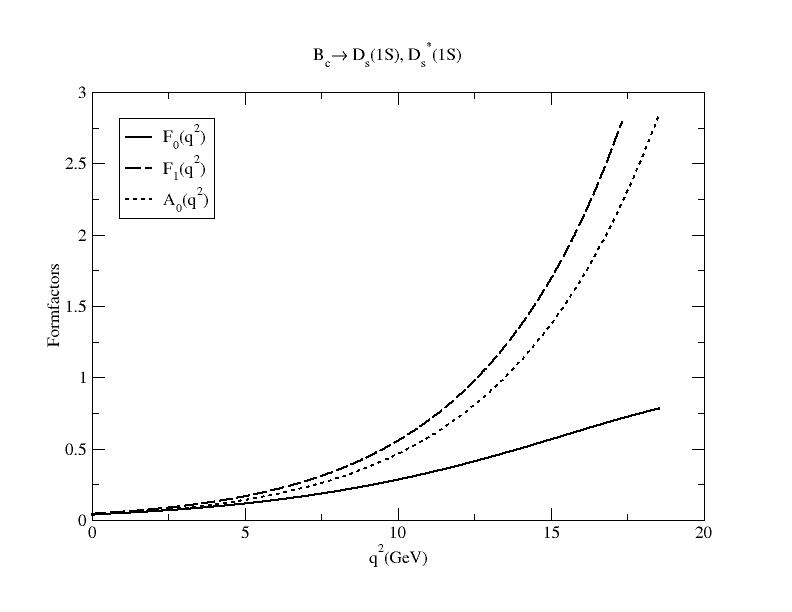}\caption{$q^2$-dependence of the form factors in the nonleptonic $B_c$-decays to final meson in 1S state.}
	\label{1S formfactors}
\end{figure}

\begin{figure}[!hbt]
	\includegraphics[width=.3\textwidth]{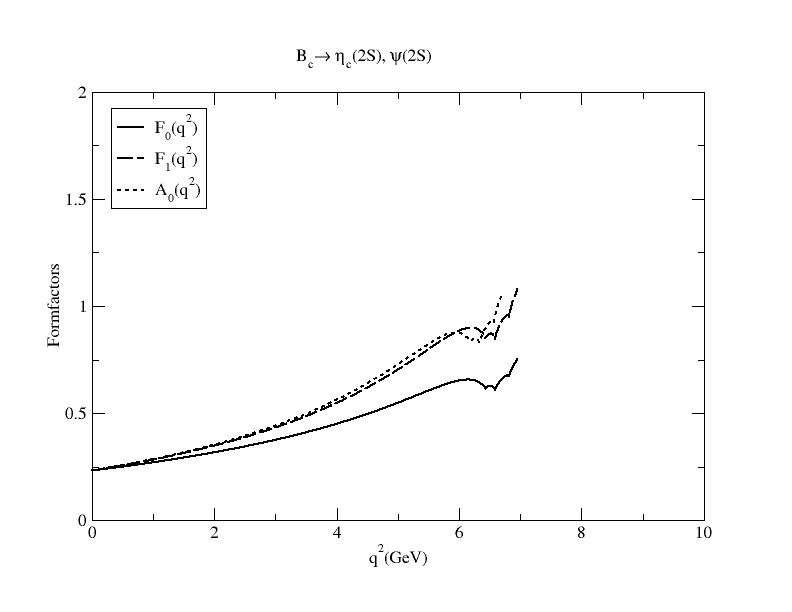}
	\includegraphics[width=.3\textwidth]{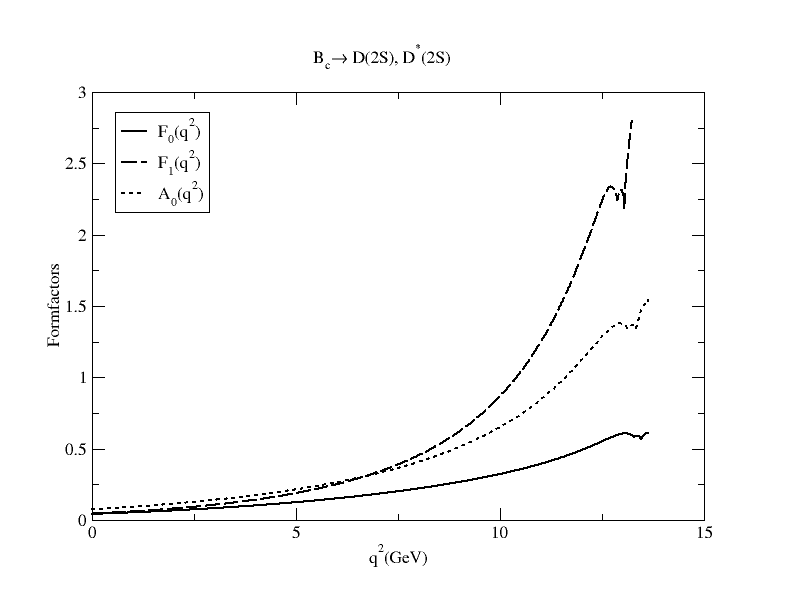}
	\includegraphics[width=.3\textwidth]{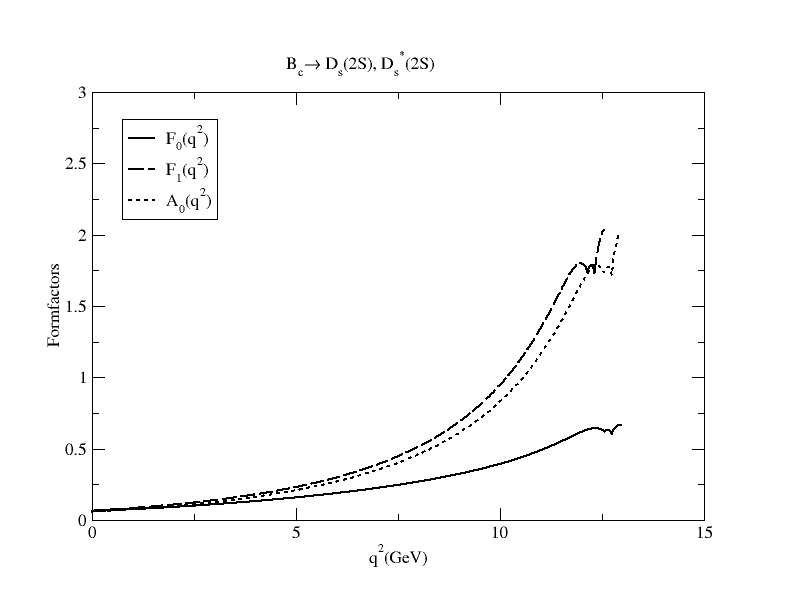}\caption{$q^2$-dependence of the form factors in the nonleptonic $B_c$-decays to final meson 2S state.}
	\label{2S formfactors}
\end{figure}

\begin{figure}[!hbt]
	\includegraphics[width=.3\textwidth]{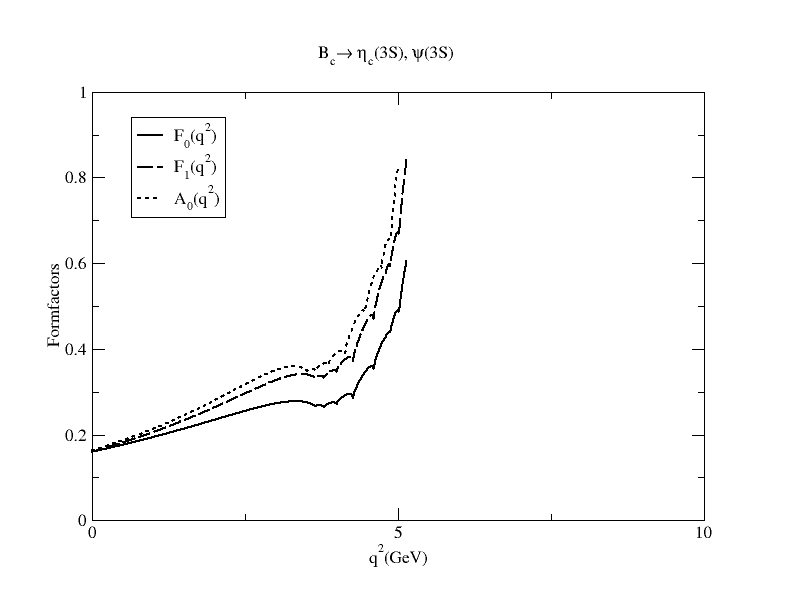}
	\includegraphics[width=.3\textwidth]{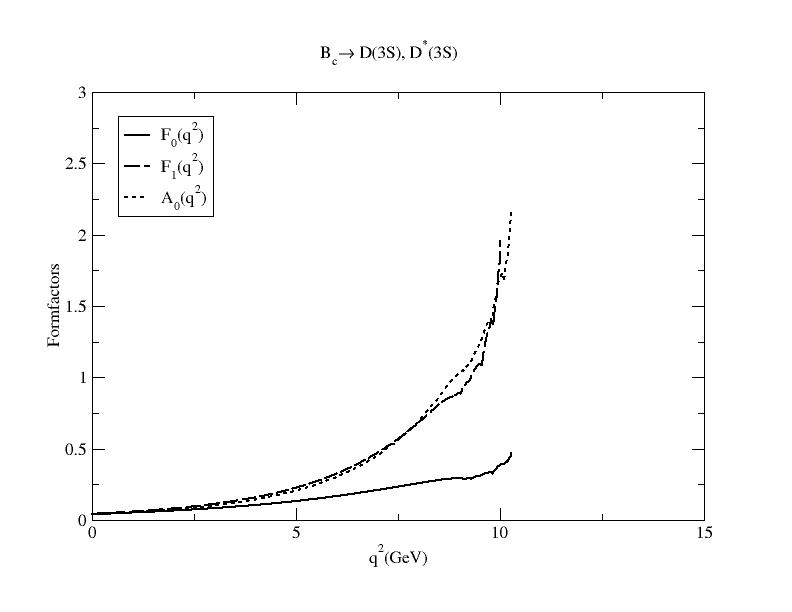}
	\includegraphics[width=.3\textwidth]{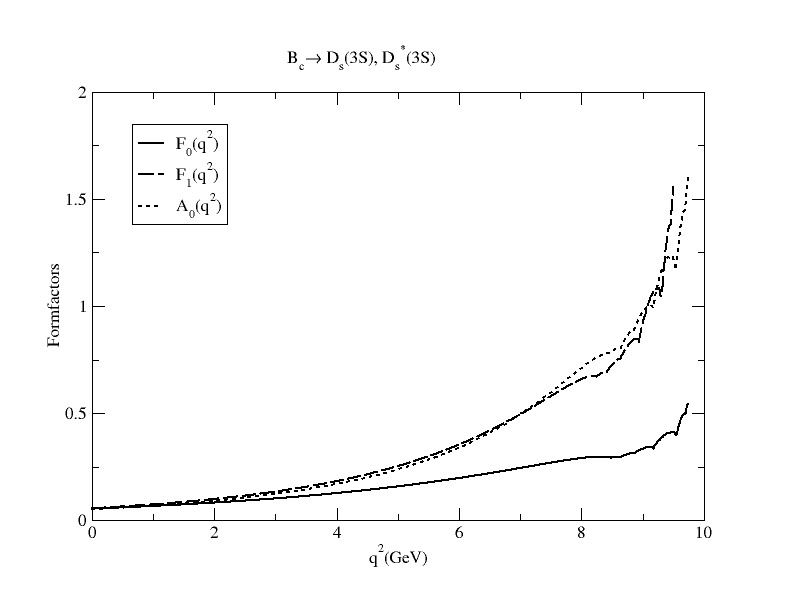}\caption{$q^2$-dependence of the form factors in the nonleptonic $B_c$-decays to final meson 3S state.}
	\label{3S formfactors}
\end{figure}

With the input parameters (41,42) we first study the $q^2$-dependence of the form factors in the allowed kinematic range: $0<q^2\le q^2_{max}$ from analytic expressions(28,32,38). In a self-consistent dynamic approach we extract the form factor from the overlapping integrals of meson wave functions where-in the $q^2$-dependence is automatically encoded in relevant expressions. This is in contrast to some model approaches cited in the literature where the form factors are determined only at one kinematic point, i.e. at $q^2\to 0$ or $q^2\to q^2_{max}$, and then extrapolated to the entire kinematic range using some phenomenological ansatz (mainly dipole or Gaussian form). Our predicted $q^2$-dependence of the form factors $F_0(q^2), F_1(q^2)$ and $A_0(q^2)$ for nonleptonic $B_c$-decays to ground and radially excited meson states are depicted in Fig. (2-4). We find that the form factors relevant to the transitions to $1S$ meson states increase with increasing $q^2$ in the entire kinematic range. This behaviour, however, is not universal as the $q^2$-dependence pattern is found different for transitions to radially excited $2S$ and $3S$ meson states. We also find that the form factors: $F_1(q^2)$ and $A_0(q^2)$ dominate $F_0(q^2)$ through out the kinematic range for all transitions. In transitions to the higher excited ($2S$ and $3S$) states, the plots for $F_1(q^2)$ and $A_0(q^2)$ almost overlap althrough above $F_0(q^2)$. Our predicted form factors at maximum recoil point: $q^2\to 0$, as listed in Table III also satisfy the requirement for pole cancellation for $B_c\to P$ transitions.

\begin{table}[!hbt]
	\renewcommand{\arraystretch}{1}
	\centering
	\setlength\tabcolsep{5pt}
	\caption{The form factors ($F_0,F_1,A_0$) at $q^2=0$ evaluated in the RIQ model for exclusive nonleptonic $B_c$-decays to 1S, 2S, 3S final state mesons.}
	\label{tab3}
	\begin{tabular}{|ccccc|}
		
		\hline
		\hline Form factors &\ \ Final meson state&\ \  ${B_c\to \eta_c, J/\psi} $\ \ &\ \  ${B_c\to D, D^*} $\ \ &\ \  ${B_c\to D_s, D_s^*} $\ \ \\
		
		\hline$ F_0 $ &&0.3057&0.0163&0.040\\
		$ F_1 $&\ \ 1S&0.3058&0.018&0.045\\
		$ A_0 $&&0.3164&0.058&0.038\\
		\hline$ F_0 $ &&0.232&0.044&0.063\\
		$ F_1 $&\ \ 2S&0.232&0.046&0.066\\
		$ A_0 $&&0.232&0.076&0.060\\
		\hline$ F_0 $ &&0.160&0.043&0.054\\
		$ F_1 $&\ \ 3S&0.160&0.045&0.056\\
		$ A_0 $&&0.163&0.0405&0.051\\
		\hline
		\hline
	\end{tabular}
\end{table}

 Before evaluating the decay rates/branching fractions it is interesting to plot the radial quark momentum distribution amplitude $|\vec{p}_b|{\cal G}_{B_c}(\vec{p}_b,-\vec{p}_b)$ for decaying $B_c$-meson at rest in its ground state along with that for the daughter mesons in their ground as well as radially excited ($2S$ and $3S$) states over allowed physical range of the quark momentum. From the plots shown in Fig.(5,6,7), we find that the overlap region in the transition to ground(1S) state is maximum. However, the overlap of momentum distribution amplitudes are found to decrease as one considers transitions to higher excited $2S$ and $3S$ states. Since invariant form factors are extracted from overlapping integrals of participating meson wave functions, one expects the form factor contribution to the decay rate/branching fractions in the decreasing order of magnitude for transition to daughter mesons from their ground to higher excited states.

\begin{figure}[!hbt]
	\includegraphics[width=.4\textwidth]{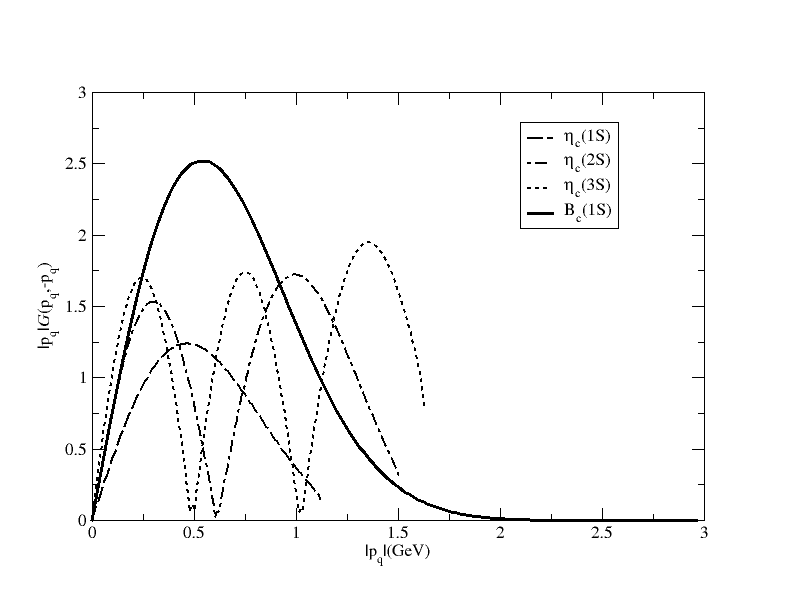}
	\includegraphics[width=.4\textwidth]{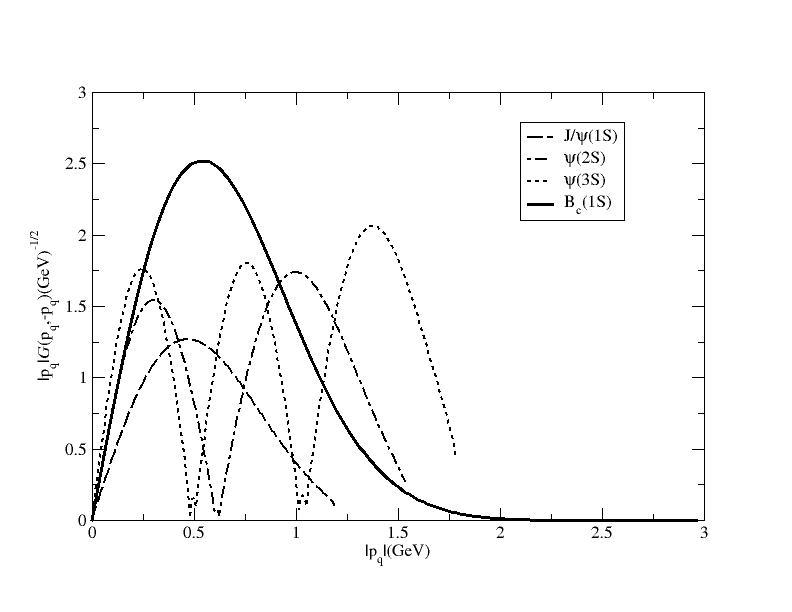}\caption{Overlap of momentum distribution amplitudes of the initial and final meson state.}
\end{figure}
\begin{figure}[!hbt]
	\includegraphics[width=.4\textwidth]{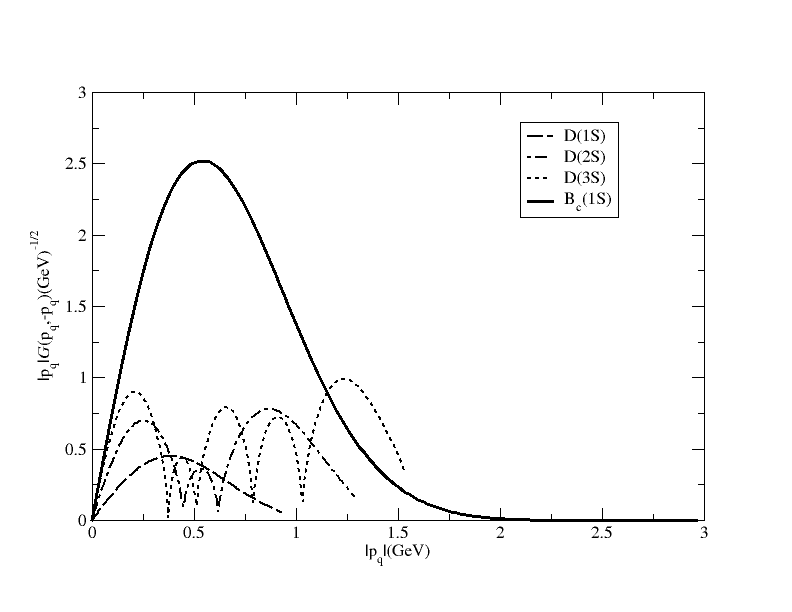}
	\includegraphics[width=.4\textwidth]{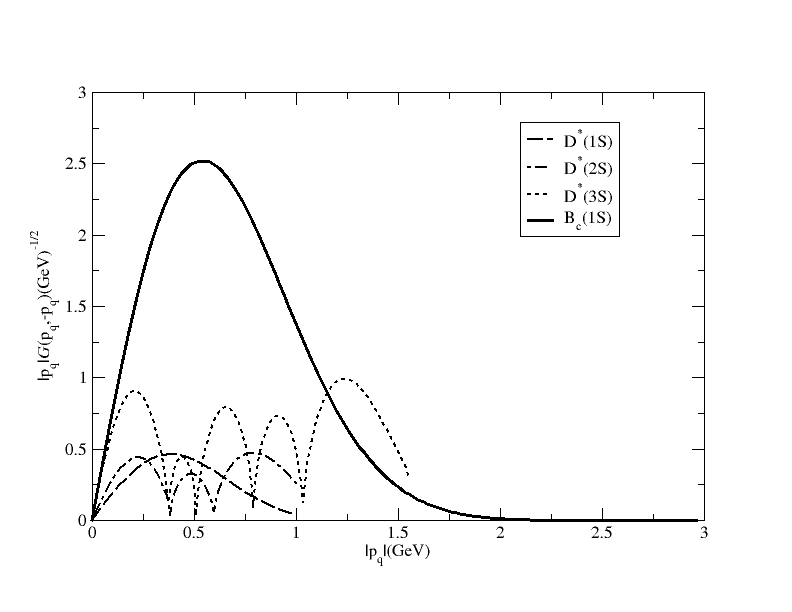}\caption{Overlap of momentum distribution amplitudes of the initial and final meson state.}
\end{figure}

\begin{figure}[!hbt]
	\includegraphics[width=.4\textwidth]{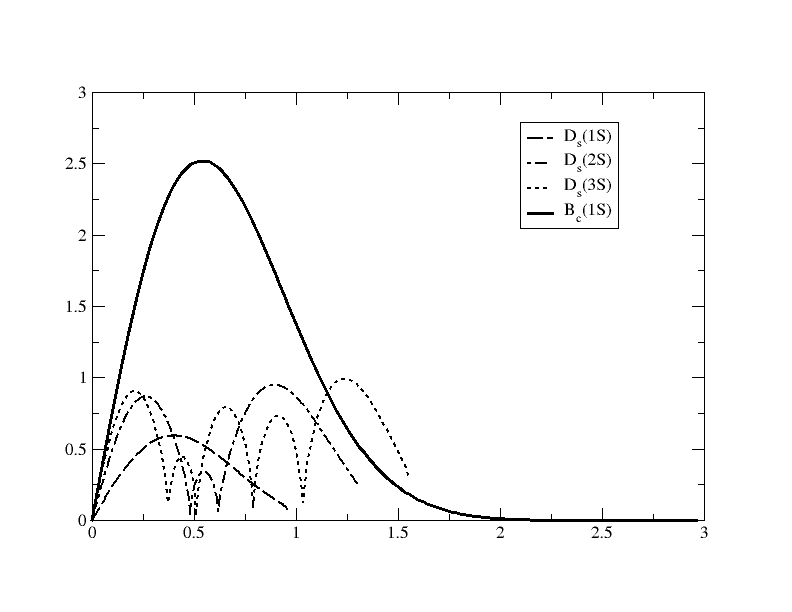}
	\includegraphics[width=.4\textwidth]{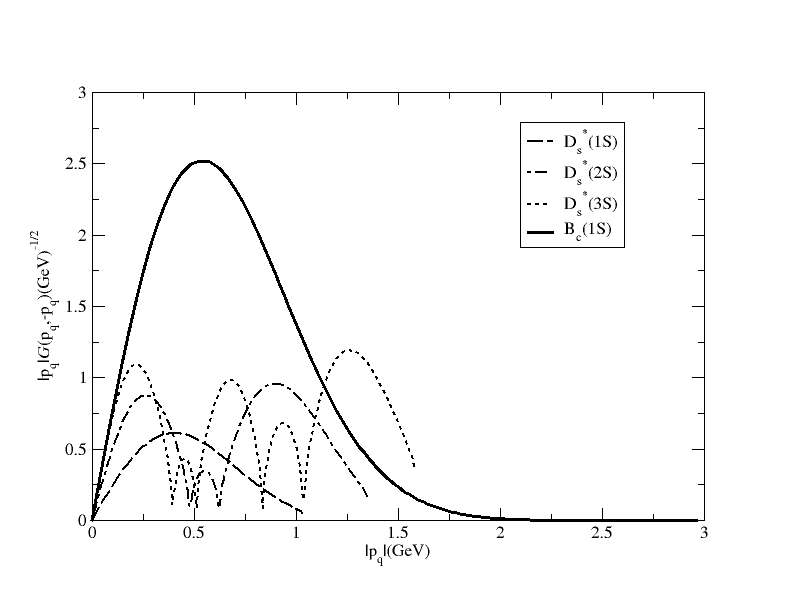}\caption{Overlap of momentum distribution amplitudes of the initial and final meson state.}
\end{figure}
\newpage
We then evaluate the decay rates from the expressions in Eq (29), (33), (39) and our results for general values of QCD coefficients $(a_1,a_2)$ of the operator product expansion are listed in Table IV to facilitate a comparison with other dynamical model predictions.

\begin{table}[!hbt]
	\renewcommand{\arraystretch}{1}
	\centering
	\setlength\tabcolsep{0.5pt}
	\caption{ Decay widths in units of $10^{-15}$ GeV for general values of the Wilson coefficients: $(a_1,a_2)$. For the sake of brevity we use the notation: $\eta_c^{'},\psi^{'}=\eta_c(2S),\psi(2S)$; $\eta_c^{''},\psi^{''}=\eta_c(3S),\psi(3S)$}
	\label{tab4}
	\begin{tabular}{|cc|cc|cc|}
		\hline Decay& Width&Decay& Width&Decay& Width\\
		\hline
		\hline ${B^-_c}\to \eta_c\pi $&0.394$a_1^{2}$& ${B^-_c}\to \eta_c^{'}\pi$&0.143$a_1^{2}$& ${B^-_c}\to \eta_c^{''}$$\pi$&0.0485$a_1^{2}$\\
		
		${B^-_c}\to \eta_cK $&0.0312$a_1^{2}$& ${B^-_c}\to \eta_c^{'}K$&0.0116$a_1^{2}$& ${B^-_c}\to \eta_c^{''}K$&0.0039$a_1^{2}$\\
		
		${B^-_c}\to \eta_cD $&$(0.306a_1+0.221a_2)^2$& ${B^-_c}\to \eta_c^{'}D$&$(0.199a_1+0.3012a_2)^2$& ${B^-_c}\to \eta_c^{''}D$&$(0.1009a_1+0.2325a_2)^2$\\
		
		${B^-_c}\to \eta_cD_s $&$(1.727a_1+1.478a_2)^2$& ${B^-_c}\to \eta_c^{'}D_s$&$(1.135a_1+1.5558a_2)^2$& ${B^-_c}\to \eta_c^{''}D_s$&$(0.512a_1+0.9910a_2)^2$\\
		\hline
		\hline 
		${B^-_c}\to \eta_c\rho $&1.237$a_1^{2}$& ${B^-_c}\to \eta_c^{'}\rho$&0.450$a_1^{2}$& ${B^-_c}\to \eta_c^{''}\rho$&0.1541$a_1^{2}$\\
		
		${B^-_c}\to \eta_cK^* $&0.0652$a_1^{2}$& ${B^-_c}\to \eta_c^{'}K^*$&0.0.0237$a_1^{2}$& ${B^-_c}\to \eta_c^{''}K^*$&0.0081$a_1^{2}$\\
		
		${B^-_c}\to \eta_cD^* $&$(0.346a_1+0.2819a_2)^2$& ${B^-_c}\to \eta_c^{'}D^*$&$(0.199a_1+0.3611a_2)^2$& ${B^-_c}\to \eta_c^{''}D^*$&$(0.0662a_1+0.2059a_2)^2$\\
		
		${B^-_c}\to \eta_cD_s^* $&$(1.893a_1+1.720a_2)^2$& ${B^-_c}\to \eta_c^{'}D_s^*$&$(1.062a_1+1.6054a_2)^2$& ${B^-_c}\to \eta_c^{''}D_s^*$&$(0.313a_1+0.558a_2)^2$\\
		\hline
		\hline
		
		${B^-_c}\to J/\psi\pi$&0.4949$a_1^{2}$&${B^-_c}\to \psi^{'}\pi$&0.173$a_1^{2}$&${B^-_c}\to \psi{''}\pi$&0.061$a_1^{2}$\\
		
		${B^-_c}\to J/\psi K$&0.0387$a_1^{2}$&${B^-_c}\to \psi^{'}K$&0.0136$a_1^{2}$&${B^-_c}\to \psi{''}K$&0.0045$a_1^{2}$\\
		
		${B^-_c}\to J/\psi D$&$(0.306a_1+0.4784a_2)^2$&${B^-_c}\to \psi^{'}D$&$(0.178a_1+0.399a_2)^2$&${B^-_c}\to \psi{''}D$&$(0.0753a_1+0.2567a_2)^2$\\
		
		${B^-_c}\to J/\psi D_s$&$(1.697a_1+2.0503a_2)^2$&${B^-_c}\to \psi^{'}D_s$&$(0.9708a_1+1.964a_2)^2$&${B^-_c}\to \psi{''}D_s$&$(0.345a_1+0.3634a_2)^2$\\
		\hline
	\end{tabular}
\end{table}

 Our predicted branching fractions for various tree-level two body nonleptonic $B_c$-decays to 1S, 2S and 3S charmonium states in comparison with other model predictions are listed in Table V, VI and VII, respectively.
 
  \begin{table}[!hbt]
 	\renewcommand{\arraystretch}{1}
 	\centering
 	\setlength\tabcolsep{0.5pt}
 	\caption{Branching ratios (in $\%$) of the $B_c\to \eta_cX$ and $B_c\to J/\psi P$ decays, where $X=P,V$ in comparision with other model predictions.}
 	\label{tab5}
 	\begin{tabular}{|c|c|c|c|c|c|c|c|c|c|c|c|}
 		\hline Decay &This work&\cite{A20} &\cite{A22}&\cite{A25}&\cite{A29}&\cite{A30}&\cite{A31}&\cite{A32}&\cite{A33}&\cite{A34}&\cite{A35} \\

 		\hline ${B^-_c}\to \eta_c\pi$& 0.0383(0.0397) &-&0.083&0.025&0.094&0.13&0.14&0.20&0.18&0.189&0.19\\
 		${B^-_c}\to \eta_c K$&0.003(0.0031)&-&0.006&0.002&0.0075&0.013&0.011&0.013&0.014&0.015&0.015\\
 		${B^-_c}\to \eta_cD^-$&0.006(0.0068)&-&-&0.005&0.014&0.010&0.014&0.015&0.0012&-&0.019\\
 		${B^-_c}\to \eta_cD_s$&0.1861(0.2169)&-&-&0.50&0.44&0.35&0.26&0.28&0.054&-&0.44\\
 		
 		\hline 
 		${B^-_c}\to \eta_c\rho$&0.120(0.124)&-&0.20&0.067&0.24&0.30&0.33&0.42&0.49&0.518&0.45\\
 		${B^-_c}\to \eta_cK^*$&0.0063(0.0065)&&0.011&0.004&0.013&0.021&0.018&0.02&0.025&0.029&0.025\\
 		${B^-_c}\to \eta_cD^*$&0.0076(0.0088)&-&-&0.003&0.012&0.0055&0.013&0.010&0.0010&-&0.019\\
 		${B^-_c}\to \eta_cD_s^*$&0.216(0.254)&-&-&0.057&0.24&0.36&0.24&0.27&0.044&-&0.37\\

 		\hline ${B^-_c}\to J/\psi\pi$&0.038(0.039)&0.0664&0.06&0.13&0.076&0.073&0.11&0.13&0.18&0.101&0.17\\
 		${B^-_c}\to J/\psi K$&0.003(0.0031)&0.00527&0.005&0.007&0.006&0.007&0.008&0.011&0.014&0.008 &0.013\\
 		${B^-_c}\to J/\psi D$&0.00369(0.00496)&0.00552&-&0.013&0.0083&0.0044&0.009&0.009&0.0009&-&0.015\\
 		${B^-_c}\to J/\psi D_s$&0.1449(0.1801)&0.137&-&0.35&0.24&0.12&0.15&0.041&0.34&-&0.34\\
 		
 		\hline
 	\end{tabular}
 \end{table}

  Our results for decays to 1S, 2S and 3S final states corresponding to both the sets of QCD parameters: one with respect to Set 1. and other with respect to Set 2. in the parenthesis are listed in the second column of each table. As expected our predicted branching fractions are obtained in the following hierarchy:
\begin{equation}
{\cal B}(B_c\to X_{c\bar{c}}(3S)M)<{\cal B}(B_c\to X_{c\bar{c}}(2S)M)<{\cal B}(B_c\to X_{c\bar{c}}(1S)M)\nonumber
\end{equation}
Our results for transitions to $2S$ and $3S$ final states are found about two and three orders of magnitude smaller, respectively, than those for $1S$ final states. The node structure of the $2S$ wave function is responsible for small branching fractions. In the calculation of the overlap integral of meson wave functions, as there is no node for the initial wave functions, the positive and negative part of the final wave function give contributions which cancel each other out yielding a small branching fraction. With regard to $3S$ final states, there are even more severe cancellations, which leads to still smaller branching fractions. As expected the tighter phase space and weaker $q^2$-dependence of form factors also lead to smaller branching fraction for transitions to higher excited $2S$ and $3S$ final states.

\begin{table}[!hbt]
	\renewcommand{\arraystretch}{1}
	\centering
	\setlength\tabcolsep{0.5pt}
	\caption{Branching ratios (in $10^{-4}$) of the $B_c^-\to \eta_c^{'}X$ and $B_c^-\to\psi^{'}P$ decays, where $X=P,V$ in comparision with other model predictions.}
	\label{tab6}
	\begin{tabular}{|c|c|c|c|c|c|c|c|c|c|}
		\hline Decay &This work&\cite{A33} &\cite{A20}&\cite{A21}&\cite{A22}&\cite{A23}&\cite{A25}&\cite{A27,A28} &\cite{A19}\\

		\hline ${B^-_c}\to \eta_c^{'}\pi$& 1.39(1.44)&2.4&-&2.4&1.7&2.2&0.66&1.67&10.3\\
		${B^-_c}\to \eta_c^{'}K$&0.11(0.117)&0.18&-&0.18&0.125&0.16&4.9$\times 10^{-2}$&0.119&-\\
		${B^-_c}\to \eta_c^{'}D$&0.161(0.21)&0.20&-&0.057&-&-&2.2$\times10^-2$&3.19$\times10^-2$&-\\
		${B^-_c}\to \eta_c^{'}D_s$&5.81(7.4)&8.7&-&0.67&-&-&0.785&4.47&-\\

		\hline ${B^-_c}\to \eta_c^{'}\rho$&12.03(12.46)&5.5&-&5.5&3.6&5.25&1.4&3.56&-\\
		${B^-_c}\to \eta_c^{'}K^*$&0.23(0.239)&0.28&-&0.26&0.15&0.25&7.15$\times10^{-2}$&0.191&-\\
		${B^-_c}\to \eta_c^{'}D^*$&0.12(0.185)&0.11&-&0.21&-&-&7.8$\times10^{-4}$&0.285&\\
		${B^-_c}\to \eta_c^{'}D_s^*$&4.61(6.13)&4.4&-&4.5&-&-&0.20&3.56&-\\

		\hline ${B^-_c}\to \psi^{'}\pi$&1.34(1.39)&2.2&2.97&3.7&1.1&0.63&2.0&1.42&6.7\\
		${B^-_c}\to \psi^{'} K$&0.105(0.109)&0.16&0.23&0.29&8$\times10^{-2}$&4.45$\times10^{-2}$&8.9$\times 10^{-2}$&0.102&-\\
		${B^-_c}\to \psi^{'} D$&0.07(0.11)&0.11&0.138&0.24&-&-&7.3$\times 10^{-2}$&1.55$\times10^{-2}$&\\
		${B^-_c}\to \psi^{'} D_s$&2.57(3.94)&4.4&3.08&5.25&-&-&1.2&2.69&-\\
		
		\hline
	\end{tabular}
\end{table}

 Our prediction for $B_c \to X_{c\bar{c}}(1S)M$ decays shown in Table V are in good agreement with the model predictions of Ref. \cite{A20,A22,A25,A29} and those for class III transitions like $B_c^-\to \eta_cD_s^-, B_c^-\to \eta_cD_s^*$ and $B_c^-\to J/\psi D_s^*$ moderately agree with those of Ref \cite{A30,A31,A32,A33,A34,A35}. For other decay modes our results are about one order of magnitude smaller than theirs. For $B_c\to X_{c\bar{c}}(2S)M$ decays, our results as listed in Table VI also agree well with those of Ref.\cite{A20,A21,A23} and agree moderately with the results of Ref.\cite{A25,A27,A28}. Compared to the branching fractions for $B_c\to \eta_c^{'}\pi$ and $B_c\to \psi^{'}\pi$ predicted in perturbative QCD approach based on $k_T$ factorization\cite{A19}, our results are found one order of magnitude smaller. Our predictions of $B_c\to X_{c\bar{c}}(3S)M$ decays shown in Table VII are found comparable to those of the model calculation based on improved Bathe-Salpetere approach \cite{A27,A28} except for $B_c \to \psi^{''}D$ decay which is found one order of magnitude larger than theirs. The dominant decay modes to $1S$ meson states are $B_c\to \eta_c(\rho^-,D_s^-,D_s^{*-})$,and $B_c\to J/\psi D_s^{-}$ which should be accessible at LHCb. The predicted branching fractions of decays to $2S$ and $3S$ states: $B_c\to \eta_c^{'}(\rho^-,D_s^-,D_s^{*-})$, $B_c\to \psi^{'}D_s^{-}$  and $B_c\to \eta_c^{''}(\rho^-,D_s^-,D_s^{*-})$, $B_c\to \psi^{''}D_s^{-}$ upto $\sim10^{-4}$ may be accessible at high luminosity hadron colliders in near future. All other decays discussed in the present work can not reach the detection ability of current experiments.
\begin{table}[!hbt]
	\renewcommand{\arraystretch}{1}
	\centering
	\setlength\tabcolsep{0.5pt}
	\caption{Predicted branching fraction (in $10^{-5}$) of the $B_c\to \eta_c^{''}x$ and $B_c\to \psi^{''}P$ where $X=P,V$.}
	\label{tab7}
	\begin{tabular}{|c|c|c|c|}
		\hline Decay &This work&\cite{A27} &\cite{A28}\\
		\hline ${B^-_c}\to \eta_c^{''}\pi$& 4.7(4.8)&2.16&-\\
		${B^-_c}\to \eta_c^{''}K$&0.38(0.39)&0.153&-\\
		${B^-_c}\to \eta_c^{''}D$&0.21(0.36)&-&-\\
		${B^-_c}\to \eta_c^{''}D_s$&7.72(11.5)&-&-\\

		\hline ${B^-_c}\to \eta_c^{''}\rho$&14.9(15.5)&4.29&-\\
		${B^-_c}\to \eta_c^{''}K^*$&0.79(0.81)&0.225&-\\
		${B^-_c}\to \eta_c^{''}D^*$&0.032(0.09)&-&-\\
		${B^-_c}\to \eta_c^{''}D_s^*$&3.2(4.6)&-&-\\

		\hline ${B^-_c}\to \psi^{''}\pi$&4.7(4.8)&3.11&-\\
		${B^-_c}\to \psi^{''}K$&0.35(0.36)&0.214&-\\
		${B^-_c}\to \psi^{''}D$&0.02(0.092)&-&3.67$\times10^{-3}$\\
		${B^-_c}\to \psi^{''}D_s$&6.6(7.9)&-&3.76\\

		\hline
	\end{tabular}
\end{table}

It is of interest to estimate the ratios of branching fractions of pairs of modes $B_c\to PV$, $B_C\to PP$ and $B_c\to VP$, $B_c\to PP$ which are expected to be $\approx 3$ from naive spin counting. We find that only for the pair $B_c^-\to \eta_c(nS)\rho^-$ and $B_c\to \eta_c(nS)\pi^-$, the naive spin counting holds good. However, for all other pairs we obtain approximate equality except for the pairs: $B_c^-\to \psi(nS)D_s^-$ and $B_c^-\to \eta_c(nS)D_s^- $ where our predicted ratio shows an inversion of spin counting ratio. The deviation of the naive spin counting is a common feature of all model predictions.\\
 
The relative size of branching fractions for nonleptonic $B_c$-decays is broadly estimated from power counting of QCD factors in the Welfenstein parameterization \cite{A66}. Accordingly the class I decay modes determined by the QCD factor $a_1$ are found to have comparatively large branching fractions which should be measured experimentally. In class III decays, which are characterized by Pauli interference, the branching fractions are determined by the relative value of $a_1$ with respect to $a_2$. Considering the positive value of $a_1=1.12$ and negative value of $a_2=-0.26$ in Set 1, for example, used in the literature, it leads to a destructive interference. As a results, the decay modes are found suppressed in comparison with the cases where interference is switched off. However, at the qualitative level, it is known that the ratio $a_2/a_1$ is a function of running coupling constant $\alpha_s$ evaluated at the factorization scale, which is shown to be positive in case of $b$-flavored meson decays corresponding to small coupling \cite{A51}. The experimental data also favour constructive interference of color-favored and color-suppressed $B_c$-decay modes. Considering positive value: $a_2^b=0.26$, our predicted branching fractions for class III decays to $1S$ and $2S$ final states find enhancement by a factor$\sim \ 2\ to\ 4$ and $\sim \ 3\ to\ 7$, respectively over that obtained with negative value of $a_2^b=-0.26$. For decays to 3S final states, the enhancement is still more significant.\\

It is possible to study the effect of Pauli interference in class III $B_c$-decays by casting the decay width in the form $\Gamma=\Gamma_0+\Delta\Gamma$, where $\Gamma_0=x_1^2a_1^2+x_2^2a_2^2$, $\Delta\Gamma=2x_1x_2a_1a_2$ and computing $\frac{\Delta\Gamma}{\Gamma_0}$ in $\% $ as done in \cite{A44,A46,A52}. The absolute values of $\frac{\Delta \Gamma}{\Gamma_0}$ for $B_c$-decays to $1S$ final states are found in the range $(32.7-63.9)\%$. Those for decay modes to $2S$ and $3S$ final states are obtained in range $(57.7-82.6)\%$ and $(46.2-97.3)\%$, respectively. Thus, we find that interference is more significant in class III $B_c$-decays to higher excited $2S$ and $3S$ states compared to those estimated for decays to $1S$ final states.

\begin{table}[!hbt]
	\renewcommand{\arraystretch}{1}
	\centering
	\setlength\tabcolsep{0.5pt}
	
	\label{tab8}
	\caption{Ratios of branching fractions: ${\cal R}$}
	\begin{tabular}{|c|c|c|}
		\hline Ratios ${\cal R}$& Our work& Experiment \\
		\hline & &$0.069\pm 0.019(Stat.)\pm 0.005(syst.)$[13] \\
		$	{\cal R}_{K/\pi}=\frac{{\cal B}(B_c\to J/\psi K)}{{\cal B}(B_c\to J/\psi \pi)}$&0.0783 & \\
		&&$0.079\pm 0.007(stat.)\pm 0.003(syst.)$[14]\\
		
			\hline &&\\
		${\cal R}_{D_s/\pi}=\frac{{\cal B}(B_c\to J/\psi D_s)}{{\cal B}(B_c\to J/\psi \pi)}$&3.7832&$2.9\pm0.57(stat.)\pm 0.003(syst.)$[6]\\
		&&\\
		
		\hline&&\\
		${\cal R}_{\psi(2S)/{J/\psi}}=\frac{{\cal B}(B_c\to \psi(2S)\pi)}{{\cal B}(B_c\to J/\psi \pi)}$&0.3394&$0.250\pm 0.068(Stat.)\pm 0.014(Syst.)\pm 0.006$[15]\\
		&&\\
		
		\hline &&\\
		${\cal R}_{\pi^+/{\mu^+\nu}}=\frac{{\cal B}(B_c\to J/\psi\ \pi)}{{\cal B}(B_c\to J/\psi \mu^+\nu_\mu)}$&0.0142&$0.049\pm 0.0028(Stat.)\pm 0.0046(Syst.)$[16]\\
		&&\\
		\hline
	\end{tabular}
\end{table}
\newpage
Finally we calculate the studied observable ${\cal R}$: ratios of branching fractions of the nonleptonic $B_c$-meson decays. It may be noted that the CKM matrix elements and decay constants do not contribute in calculating the ratios ${\cal R}$. The QCD parameter which appears in the decay amplitudes and the theoretical uncertainties caused by naive factorization for nonleptonic decays also get  cancelled a lot in calculating the observable $\cal R$. Contrary to other observables, the above mentioned ratios, in which the production of $B_c$-meson is cancelled totally, provide an essential test of the decays. Besides giving useful information about the form factors, these ratios could offer a test of the adopted quark model. Our predicted observables $(\cal R)$ are listed in Table VIII. One can see that our results for the ratios: ${\cal R}_{K/\pi},{\cal R}_{{D_s}/\pi}$ and ${\cal R}_{{\psi(2S)}/{J/\psi}}$ are consistent with the LHCb  data within experimental uncertainties. Only the predicted ratio ${\cal R}_{\pi/{\mu\nu}}$ is found to be underestimated. Since the related ratios are mostly determined by hadron transition, the agreement between the observed values and our predictions indicates a strong support to our approach to study nonleptonic $B_c$-decays within the framework of the RIQ model.

\section{Summary and conclusion}
In this work we study the exclusive two-body nonleptonic $B_c\to X_{c\bar{c}}M$ decays, where $X_{c\bar{c}}$ is a S-wave charmonium state and $M$ is either a pseudoscalar $(P)$ or a vector $(V)$ meson. We consider here three categories of decays: $B_c\to PP, PV, VP$ decays within factorization approximation in the framework of the relativistic independent quark(RIQ) model based on a flavor-independent interaction potential in the scalar-vector harmonic form. The weak decay form factors representing decay amplitudes and their $q^2$-dependence are extracted in the entire kinematic range of $0\le q^2\le q^2_{max}$ from the overlapping integrals of the meson wave functions obtainable in the RIQ model.\\

In calculating the decay modes, we find our predicted branching fractions in a wide range $\sim 10^{-2}$ to $10^{-5}$, in reasonable agreement with most other model predictions. The dominant decay modes to 1S charmonium states are $B_c^-\to \eta_c (\rho^-,D_s^-,D_s^{*-})$ and $B_c\to J/\psi D_s^-$ which should be experimentally accessible. The branching fractions to 2S and 3S charmonium states: $B_c^-\to \eta_c^{'} (\rho^-,D_s^-,D_s^{*-})$, $B_c\to \psi^{'} D_s^-$ and $B_c\to \eta_c^{''} (\rho^-,D_s^-,D_s^{*-})$, $B_c\to \psi^{''} D_s^-$ predicted upto $\sim 10^{-4}$ may be accessible at high luminosity hadron colliders in near future. Other decay modes with lower branching fractions can not reach the detection ability of current experiments. As expected our predicted branching fractions are obtained in the hierarchy:
\begin{equation}
{\cal B}(B_c\to X_{c\bar{c}}(3S)M)< {\cal B}(B_c\to X_{c\bar{c}}(2S)M)<{\cal R}(B_c\to X_{c\bar{c}}(1S)M)\nonumber
\end{equation}
This is due to the nodal structure of the participating meson wave functions in the decays to higher excited states, tighter phase space and weaker $q^2$-dependence of the form factors for the decays to higher excited states in comparison to the decays to ground(1S) states.\\

The class I decays determined by QCD co-efficient $a_1$ are found to have comparatively large branching fractions which should be measured experimentally. The class III decays, characterized by Pauli interference, are determined by both QCD parameters $a_1$ and $a_2$. Considering the positive values of $a_1$ and negative value of $a_2$ used in the literature: it leads to a destructive interference, as a result of which, this class of decays are found to be suppressed compared to the case where the interference is switched off. We find the interference is more significant in class III $B_c$-decays to higher excited 2S and 3S states compared to that found in the decays 1S final states.\\  

In the wake of recent measurements of the ratios of branching fractions by the LHCb Collaborations for nonleptonic $B_c$-decays, these ratios are calculated in the present study and our predicted ratios of branching fractions: ${\cal R}_{K/\pi}$, ${\cal R}_{D_s/\pi}$ and ${\cal R}_{\psi(2S)/{J/\psi}}$ are consistent with the LHCb data within experimental uncertainties, although our prediction of ${\cal R}_{\pi/\mu\nu}$ is found to be underestimated. Since the CKM parameters and decay constants do not contribute, and the QCD parameter and the theoretical uncertainties due to naive factorization are also cancelled a lot in calculating the ratios ${\cal R}$, these predicted ratios, contrary to other observables could offer a test for the phenomenological models adopted in the description of nonleptonic decays. Our results for the weak form factors, branching fractions and ratios $\cal R$ etc. indicate that the approach adopted here works well to describe nonleptonic $B_c$-decays within factorization approximation in the framework of the RIQ model.

\appendix
\section{CONSTITUENT QUARK ORBITALS AND MOMENTUM PROBABILITY AMPLITUDES}\label{app}
In RIQ model a meson is picturised as a color-singlet assembly of a quark and an antiquark independently confined by an effective and average flavor independent potential in the form:
$U(r)=\frac{1}{2}(1+\gamma^0)(ar^2+V_0)$ where ($a$, $V_0$) are the potential parameters. It is believed that the zeroth order quark dynamics  generated by the phenomenological confining potential $U(r)$ taken in equally mixed scalar-vector harmonic form can provide adequate tree level description of the decay process being analyzed in this work. With the interaction potential $U(r)$ put into the zeroth order quark Lagrangian density, the ensuing Dirac equation admits static solution of positive and negative energy as: 
\begin{eqnarray}
\psi^{(+)}_{\xi}(\vec r)\;&=&\;\left(
\begin{array}{c}
\frac{ig_{\xi}(r)}{r} \\
\frac{{\vec \sigma}.{\hat r}f_{\xi}(r)}{r}
\end{array}\;\right)U_{\xi}(\hat r)
\nonumber\\
\psi^{(-)}_{\xi}(\vec r)\;&=&\;\left(
\begin{array}{c}
\frac{i({\vec \sigma}.{\hat r})f_{\xi}(r)}{r}\\
\frac{g_{\xi}(r)}{r}
\end{array}\;\right){\tilde U}_{\xi}(\hat r)
\end{eqnarray}
where, $\xi=(nlj)$ represents a set of Dirac quantum numbers specifying 
the eigen-modes;
$U_{\xi}(\hat r)$ and ${\tilde U}_{\xi}(\hat r)$
are the spin angular parts given by,
\begin{eqnarray}
U_{ljm}(\hat r) &=&\sum_{m_l,m_s}<lm_l\;{1\over{2}}m_s|
jm>Y_l^{m_l}(\hat r)\chi^{m_s}_{\frac{1}{2}}\nonumber\\
{\tilde U}_{ljm}(\hat r)&=&(-1)^{j+m-l}U_{lj-m}(\hat r)
\end{eqnarray}
With the quark binding energy $E_q$ and quark mass $m_q$
written in the form $E_q^{\prime}=(E_q-V_0/2)$,
$m_q^{\prime}=(m_q+V_0/2)$ and $\omega_q=E_q^{\prime}+m_q^{\prime}$, one 
can obtain solutions to the resulting radial equation for 
$g_{\xi}(r)$ and $f_{\xi}(r)$in the form:
\begin{eqnarray}
g_{nl}&=& N_{nl} (\frac{r}{r_{nl}})^{l+l}\exp (-r^2/2r^2_{nl})
L_{n-1}^{l+1/2}(r^2/r^2_{nl})\nonumber\\
f_{nl}&=& N_{nl} (\frac{r}{r_{nl}})^{l}\exp (-r^2/2r^2_{nl})\nonumber\\
&\times &\left[(n+l-\frac{1}{2})L_{n-1}^{l-1/2}(r^2/r^2_{nl})
+nL_n^{l-1/2}(r^2/r^2_{nl})\right ]
\end{eqnarray}
where, $r_{nl}= a\omega_{q}^{-1/4}$ is a state independent length parameter, $N_{nl}$
is an overall normalization constant given by
\begin{equation}
N^2_{nl}=\frac{4\Gamma(n)}{\Gamma(n+l+1/2)}\frac{(\omega_{nl}/r_{nl})}
{(3E_q^{\prime}+m_q^{\prime})}
\end{equation}
and
$L_{n-1}^{l+1/2}(r^2/r_{nl}^2)$ etc. are associated Laguerre polynomials. The radial solutions yields an independent quark bound-state condition in the form of a cubic equation:
\begin{equation}
\sqrt{(\omega_q/a)} (E_q^{\prime}-m_q^{\prime})=(4n+2l-1)
\end{equation}
The solution of the cubic equation provides the zeroth order binding energies of 
the confined quark and antiquark for all possible eigenmodes.

In the relativistic independent particle picture of this model, the constituent quark 
and antiquark are thought to move independently inside the $B_c$-meson bound state 
with momentum $\vec p_b$ and $\vec p_c$, respectively. Their individual momentum probability 
amplitudes are obtained in this model via momentum projection of respective quark orbitals (A1) in following forms:

For ground state mesons:($n=1$,$l=0$)
\begin{eqnarray}
G_b(\vec p_b)&&={{i\pi {\cal N}_b}\over {2\alpha _b\omega _b}}
\sqrt {{(E_{p_b}+m_b)}\over {E_{p_b}}}(E_{p_b}+E_b)\nonumber\\
&&\times\exp {(-{
		{\vec {p_b}}^2\over {4\alpha_b}})}\nonumber\\
{\tilde G}_c(\vec p_c)&&=-{{i\pi {\cal N}_c}\over {2\alpha _c\omega _c}}
\sqrt {{(E_{p_c}+m_c)}\over {E_{p_c}}}(E_{p_c}+E_c)\nonumber\\
&&\times\exp {(-{
		{\vec {p_c}}^2\over {4\alpha_c}})}
\end{eqnarray}

\noindent For the excited meson state:(n=2, l=0)
\begin{eqnarray}
G_b(\vec p_b)&&={{i\pi {\cal N}_b}\over {2\alpha _b}}
\sqrt {{(E_{p_b}+m_b)}\over {E_{p_b}}} {(E_{p_b}+E_b)\over{(E_b+m_b)}}\nonumber\\ 
&&\times({\vec {p_b}^2\over {2\alpha _b}}-{3\over 2})
\exp {(-{{\vec p_b}^2\over {4\alpha_b}})}\nonumber\\
{\tilde G}_c(\vec p_c)&&={{i\pi {\cal N}_c}\over {2\alpha _c}}
\sqrt {{(E_{p_c}+m_c)}\over {E_{p_c}}}
{(E_{p_c}+E_c)\over {(E_c+m_c)}}\nonumber\\
&&\times({\vec {p_c}^2\over {2\alpha _c}}-{3\over 2})
\exp {(-{
		{\vec p_c}^2\over {4\alpha_c}})}
\end{eqnarray}                   		
\noindent For the excited meson state (n=3, l=0)
\begin{eqnarray}  
G_b(\vec p_b)&&={{i\pi {\cal N}_b}\over {2\alpha _b}}
\sqrt {{(E_{p_b}+m_b)}\over {E_{p_b}}} {(E_{p_b}+E_b)\over{(E_b+m_b)}} \nonumber\\
&&\times({\vec {p_b}^4\over {8 {\alpha _b}^2}}-{{5{\vec p_b}^2}\over {4\alpha_b}}+{{15\over 8}})
\exp {(-{{\vec p_b}^2\over {4\alpha_b}})}\nonumber\\
{\tilde G}_c(\vec p_c)&&={{i\pi {\cal N}_c}\over {2\alpha _c}}
\sqrt {{(E_{p_c}+m_c)}\over {E_{p_c}}}
{(E_{p_c}+E_c)\over {(E_c+m_c)}} \nonumber\\
&&\times({\vec {p_c}^4\over {8 {\alpha _c}^2}}-{{5{\vec p_c}^2}\over {4\alpha_c}}+{{15\over 8}})
\exp {(-{{\vec p_c}^2\over {4\alpha_c}})}
\end{eqnarray} 
The binding energies of constituent quark and antiquark for the ground state of $B_c$ meson as well as the ground and excited final meson states for $n=1,2,3$; $l=0$ can also be obtained by solving respective cubic equations representing appropriate bound state conditions.


\newpage 


	\begin{thebibliography}{90}
		\bibitem{A1}
		PARTICLE DATA GROUP Collaboration, Review of Particle Physics,{\it Phys. Rev.} D {\bf 98} (2018).
	
		\bibitem{A2}
	 F.Abe {\it et al.}, CDF Collaboration, {\it Phys.Rev.}  D {\bf 58}, 112004 (1998);	 F.Abe {\it et al.}, CDF Collaboration, {\it Phys. Rev. Lett.}{\bf 81}, 2432 (1998).
	 
	 \bibitem{A3}
	 BEllE Collaboration, Observation of the $\eta_c(2S)$ in exclusive $B\to KK_sK^-\pi^+$ decays, {\it Phys. Rev. Lett.} {\bf 89} (2002) 102001 [hep-ex/0206002].
	 \bibitem{A4}
	 	R. Aajj {\it et al.}(LHCb Collaboration){\it Phys. Rev. Lett.} {\bf 108}, 251802 (2012).
	 	\bibitem{A5}
	 		R. Aajj {\it et al.}(LHCb Collaboration){\it JHEP} {\bf 09}, 075 (2013).
	 		
	 \bibitem{A6}		  
	 	R. Aajj {\it et al.}(LHCb Collaboration){\it Phys. Rev.} D {\bf 87}, 112012 (2013). 
	 
	 	\bibitem{A7}
	 R. Aajj {\it et al.}(LHCb Collaboration){\it JHEP} {\bf 11}, 094 (2013).
	 
	 \bibitem{A8}
	 R. Aajj {\it et al.}(LHCb Collaboration){\it Phys. Rev. Lett.} {\bf 111}, 181801 (2013).
	
	\bibitem{A9}
	N. Brambilla {\it et al.} (Quarkonium Working Group), arXive:hep-ph/0412158.
	
	\bibitem{A10}
	I. P. Gouz, {\it et al.}, Yad Fiz. {\bf 67}, 1581 (2004)[{\it Phys. At. Nucl.} {\bf 67}, 1559 (2004)].
	
	\bibitem{A11}
	S. Descotes-Genon, J. He, E. Kou and P. Robbe, {\it Phys. Rev.} D {\bf 80}, 114031(2009); C-H. Chang and X.-G. Wu, {\it Eur. Phys. J.} C {\bf 38}, 267(2004).
	
	\bibitem{A12}
	Y. N. Gao {\it et al.}, {it Chin. Phys. Lett.} {\bf 27}, 061302 (2010).
	
	\bibitem{A13}
	R. Aajj {\it et al.}(LHCb Collaboration){\it First observation of the decay $B_c\to J/\psi K^+$, JHEP} {\bf 1309}, (2013) 075, arXive:1306-6723[hep-ph].
	
	\bibitem{A14}
	R. Aaij {\it et al.}, (LHCb Collaboration) {\it Measurement of the ratio of branching fractions $\frac{{\cal B}(B_C\to J/\psi K^+)}{{\cal B}(B_C\to J/\psi \pi^+)}$ JHEP} {\bf 1609} (2016) 153, arXive:1607.06823[hep-ph].
	
	\bibitem{A15}
		R. Aajj {\it et al.}(LHCb Collaboration){\it Phys. Rev.} D {\bf 87}, 071103 (2013).
		
	\bibitem{A16}
	R. Aajj {\it et al.}(LHCb Collaboration), {\it Measurement of the ratio of $B_c^+$ branching fractions to $J/\psi \pi^+$ and $J/\psi \mu^+\nu_{\mu}$, Phys. Rev.} D {\bf 90}(3) (2014), 032009.
	\bibitem{A17}
		X. Liu, Z-J. Xiao, and C-D. Lu, {\it Phys. Rev.} D {\bf 81}, 014022 (2010).
		\bibitem{A18}
	Zhou Rui1, Hong Li, Guang-xin Wang, Ying Xiao,{\it Eur. Phys. J. C} (2016) 76:564.
	\bibitem{A19}
	Z. Rui, W. F. Wang, G. Wang, Li-hua Song, C. D. L$\ddot{u}$, {\it Eur. Phys. J. C} (2015) 75:293.	
	\bibitem{A20}
	H. W. Ke, T. Liu, X. Q. Li, {\it Phys. Rev.} D {\bf 89}, 017501 (2014).
	\bibitem{A21}
	I. Bediaga, J. H. Mu$\tilde{n}$oz, arXiv:1102.2190v2 [hep-ph].
	\bibitem{A22}
	D. Ebert, R. Faustov and V. Galkin, {\it Weak decays of the $B_c$ meson to charmonium and D mesons in the relativistic quark model, Phys. Rev. D} {\bf 68} (2003) 094020.
	\bibitem{A23}
	J. F. Liu, K. T Chao, {\it Phys. Rev.} D {\bf 56}, 4133 (1997).
	\bibitem{A24}
	C. H. Chang, H. F. Fu, G. L. Wang, J. M. Zhang, arXiv:1411.3428; CHANG Chao Hsi, FU Hui Feng, WANG Guo Li, ZHANG Jin Mei, doi:10.1007/s11433-015-5671-x.


	\bibitem{A25}
		P. Colangelo, F. De Fazio, {\it Phys. Rev. } D {\bf 61}, 034012 (2000).
	
	
	\bibitem{A26}
	Hui-feng Fu, Yue Jiang, C. S. Kim and Guo-Li Wang, doi:10.1007/JHEP06(2011)015.
	
	 \bibitem{A27}
	 Tian Zhou, Ti anhong, Yue Jiang, Lei Huo, Guo-Li Wang, arXive:2006.05704v2[hep-ph]21 oct 2020.
	 \bibitem{A28}
	 T. Zhou, T. Wang, H. F. Fu, Z. H. Wang, L. Huo, G. Li Wang, arXiv:2012.06135v2[hep-ph] 13 April 2021.
	 \bibitem{A29}
	 E. Hernandez, J. Nieves and J. M. Verde-Velasco, {\it Phys. Rev.} D {\bf 74}, 074008(2006).
		\bibitem{A30}
	A. Y Anisimov, I. M. Narodetasky, C. Semay, and B. Silvestre-Brac, {\it Phys. Lett.} B {\bf 452}, 129 (1999); A. Y Anismov, P. Y. Kulikov, I. M. Narodetsky and K. A. Ter-Martirosian, Yad. Fiz {\bf 62},1868 (1999).
	\bibitem{A31}
	A.Abd EI-Hady, J. H. Munoz and J. P. Vary, {\it Phys. Rev.} D {\bf 62}, 014019 (2000).
	
	\bibitem{A32}
	V.V. Kiselev, A. E. Kovalsky and A. K. Likhoded, {\it Phys. At. Nucl.} {\bf 64}, 1860 (2001); {\it Nucl. Phys.} B{\bf 585}, 353(2000); V.V. Kiselev, O. N. Pakhomova and V.A. Saleev, {\it J. Phys.} G{\bf 28}, 595(2002); V. V. Kiselev, arXiv:[hep-ph]/0211021; X. Q. Yu and X. L. Zhou, {\it Phys. Rev.} D {\bf 81}, 037501 (2010).
		\bibitem{A33}
	C. H. Chang, Y. Q. Chen, {\it Phys. Rev.} D	{\bf 49}, 3399 (1994).
	
	\bibitem{A34}
	Adios Issadykov, Mikhali A. Ivanov, {\it Phys. Lett.} B {\bf 783}, 178-182 (2018).
		 
	\bibitem{A35}
	M. A. Ivanov, J. G. Korner and P. Santrorelli, {\it Phys. Rev.} D {\bf 73}, 054024 (2006).
	
		\bibitem{A36}
	N. Barik, B. K. Dash, {\it Phys. Rev.} D {\bf 33}, 1925 (1986); N. Barik, B. K. Dash, P. C. Dash {\it Pramana J. Phys.} {\bf 29} 543 (1987); N. Barik, P. C. Dash, {\it Phys. Rev.} D {\bf 47}, 2788 (1993).
	\bibitem{A37}
	N. Barik, P. C. Dash, A. R. Panda, {\it Phys. Rev.} D {\bf 46}, 3856 (1992); N. Barik, P. C. Dash, {\it Phys. Rev.} D {\bf 49}, 299 (1994); M. Priyadarsini, P. C. Dash, S. Kar, S. P. Patra, N. Barik, {\it Phys. Rev.} D {\bf 94}, 113011 (2016); N. Barik, P. C. Dash, {\it Mod. Phys. Lett.} A {\bf 10}, 103 (1995); N. Barik, S. Kar, P. C. Dash, {\it Phys.Rev.} D {\bf 57}, 405 (1998); N. Barik, Sk. Naimuddin, S. Kar, P. C. Dash, {\it Phys. Rev.} D {\bf 63}, 014024 (2000).
	
	\bibitem{A38}
	
	N. Barik, P. C. Dash, A. R. Panda, {\it Phys. Rev.} D {\bf 47}, 1001 (1993); N. Barik, P. C. Dash, {\it Phys. Rev.} D {\bf 47}, 2788 (1993); N. Barik, Sk. Naimuddin, P. C. Dash, S. Kar, {\it Phys. Rev.} D {\bf 77}, 014038 (2008); N. Barik, Sk. Naimuddin, P. C. Dash, S. Kar, {\it Phys. Rev.} D {\bf 78}, 114030 (2008); N. Barik, Sk. Naimuddin, P. C. Dash, {\it Mod. Phys.} A{\bf 24},2335 (2009).
	
	\bibitem{A39}
	N. Barik, P.C. Dash, {\it Phys. Rev.} D {\bf 53},1366 (1996);N. Barik, S. K. Tripathy, S. Kar, P. C. Dash, {\it Phys. Rev.} D {\bf 56}, 4238 (1997);	N. Barik, Sk. Naimuddin, P. C. Dash, S. Kar, {\it Phys. Rev. } D {\bf 80}, 074005 (2009).

	\bibitem{A40}
	S. Patnaik, P. C. Dash, S. Kar, S. P. Patra, N. Barik, {\it Phys. Rev.}	D {\bf 96},116010 (2017); S. Patnaik, P. C. Dash, S. Kar and N. Barik, {\it Phys. Rev.} D {\bf 97}, 056025 (2018).
	
	\bibitem{A41}
	S. Patnaik, L. Nayak, P. C. Dash, S. Kar, N. Barik, {\it Eur. Phys. J. Plus} (2020) 135:936.
	
	\bibitem{A42}
	L. Nayak, S. Patnaik, P. C. Dash, S. Kar and N. Barik {\it Phys. Rev.} D {\bf 104}, 036012 (2021).
	

	
	\bibitem{A43}
	N. Barik, S. Kar and P. C. Dash {\it Phys. Rev.}
 D {\bf 63}, 114002 (2001); N. Barik, Sk Naimuddin, P. C. Dash and S. Kar, {\it Phys. Rev.} D {\bf 80}, 014004 (2009).
 
 \bibitem{A44}
 Sk Naimuddin, S. Kar, M. Priyadarshini, N. Barik, P. C. Dash, {\it Phys. Rev.} D {\bf 86}, 094028 (2012);S. Kar, P. Dash, M. Priyadarsini, Sk Naimuddin, and N. Barik, {\it Phys. Rev.} D {\bf 88}, 094014 (2013).  
 \bibitem{A45}
 M. Wirbel, B. Stech, and M. Bauer, {\it Z. Phys.} C {\bf 29}, 637
 (1985); M. Bauer, B. Stech, and M. Wirbel, {\it Z. Phys.} C 34,
 103 (1987); L.-L. Chau, H.-Y. Cheng, W. K. Sze, H. Yao,
 and B. Tseng, {\it Phys. Rev.} D {\bf 43}, 2176 (1991); 58, 019902
 (1998).
 

	 
	 \bibitem{A46}
	 I. P. Gouz, V. V. Kiselev, A. K. Likhoded, V. I.
	 Romanovsky, and O. P. Yushchenko, {\it Phys. At. Nucl.} {\bf 67},
	 1559 (2004).
	 \bibitem{A47}
	 M. Beneka, G. Buchalla, M. Neubert and C.T. sachrajda, {\it Phys. Rev. lett.} {\bf 83}, 1914 (1999); {\it Nucl. Phys.} B {\bf 591}, 313(2000); B{\bf 606}, 245 (2001); M. Beneka and M. Neubert, {\it Nucl. Phys.} B{bf 675}, 33 (2003); C. E. Thomas, {\it Phys. Rev.} D {\bf 73}, 054016 (2006).
	 
	 
	 \bibitem{A48}
	 J. D. Bjorken, in Proceedings of International Workshop, Crete, Greece, 1988, edited by G. Branco and J. Reeo, Development in high Energy Physics[{\it Nucl. Phys.} B Proc. Suppl. 11, 325(1989)].
	 
	 \bibitem{A49}
	 A. J. Buras, J. M. Gerard and R. Ruckl, {\it Nucl. Phys. } B{\bf 268}, 16(1986).
	 \bibitem{A50}
	 M. Neubert, {\it Phys. Rep.} {\bf 245}, 259 (1994).
	 

	 \bibitem{A51}
	 M. Neubert and B.Stech, {\it Adv. Ser. Dir. High Energy Phys.} {\bf 17}, 294 (1998).
	 
	 \bibitem{A52}
	 H. M. Choi and C. R. Ji, {\it Phys. Rev. } D {\bf 80}, 114003 (2009).
	 \bibitem{A53}
	  J. Sun, D. Du, and Y. Yang, {\it Eur. Phys. J.} C {\bf 60}, 107 (2009);
	 J. Sun, Y. Yang, W. Du, and H. Ma, {\it Phys. Rev.} D {\bf 77},
	 114004 (2008); J. Sun, G. Xue, Y. Yang, G. Lu, and D. Du,
	 {\it Phys. Rev.} D {\bf 77}, 074013 (2008).
	 
	 \bibitem{A54}
	 C. E. Thomas, {\it Phys. Rev.} D {\bf 73}, 054016 (2006).
	 
	
	  
	  
	  \bibitem{A55}
	  G. Buchalla {\it et al., Eur. Phys. J.} C {\bf 57}, 309 (2008); A. J. Buras, arXiv:[hep-ph]/9806471,(2000).
	   \bibitem{A56}
	  N. Sharma and R. C. Verma, {\it Phys. Rev.} D {\bf 82}, 094014(2010); N. Sharma, R. Dhir and R. C. Verma, {\it J. Phys.} G: {Nucl. Part. Phys. } {\bf 37}, 075013(2010); R. Dhir and R. C. Verm, {\it Phys. Rev.} D {\bf 79}, 034004(2009); N. Sharma, {\it Phys. Rev.} D {\bf 81}, 014027 (2010). 
	  \bibitem{A57} 
	  B. Margolis and R. R. Mendel, {\it Phys. Rev} D {\bf 28}, 468(1983).
	  \bibitem{A58}
	  P. Zyla {\it etal.} (Particle Data Group), {\it Prog. Ther. Exp. Phys.} 2020, 083C01(2020).
	  \bibitem{A59}
	  P. Abreu {\it etal.}(DELPHI Collaboration), {Phys. Lett.} B {\bf 426}, 231 (1998).
	  
	\bibitem{A60}
	R. Aaij {\it etal.} (LHCb Collaboration), JHEP 02, 133(2016).
	\bibitem{A61}
	S. Godfrey and K. Moats, {\it Phys. Rev.} D {\bf 93}, 034035 (2016).
	
	\bibitem{A62}
	H. Wang, Z. Yan and J. Ping, {\it Eur. Phys. J.} C {\bf 75}, (2015) 196[1412.7068].
	
	\bibitem{A63}
	{\it Nucl. Phys.} B {\bf 735} (2014) 12-18.
	\bibitem{A64}
	{\it Nucl. Phys.} B {\bf 883}(2014) 306-327.
	
	\bibitem{A65}
	T. E. Browder and K. Honscheid, {\it Prog. Part. Nucl. Phys.}{\bf 35}, 81 (1995); M. Neubert, V. Rieckert, B. Stech, and Q. P. Xu, in {\it Heavy Flavors} edited by A. J. Buras and H. Lindner	(World Scientific, Singapore, 1992) and references therein.
	
	\bibitem{A66}
	L. Wolfenstein, {\it Phys. Rev. Lett.} {\bf 51}, 1945[1983].
	 
		

	\end{thebibliography}
\end{document}